\documentclass[useAMS]{mn2e}

\usepackage{graphicx,color,soul}
\usepackage{latexsym}
\usepackage{multirow}
\usepackage{amsmath,amssymb}
\usepackage{epstopdf}
\usepackage{subfig}
\usepackage[draft=false]{hyperref}

\title[Newtonian CAFE]{Newtonian CAFE: a new ideal MHD code to study the solar atmosphere}

\author[J. J. Gonz\'alez-Avil\'es, A. Cruz-Osorio, 
        F. D. Lora-Clavijo and F. S. Guzm\'an]
       {J. J. Gonz\'alez-Avil\'es$^{1}$ \thanks{E-mail:javiles@ifm.umich.mx (JJGA)},
        A. Cruz-Osorio$^{2}$ \thanks{E-mail:aosorio@astro.unam.mx (ACO)},
        F. D. Lora-Clavijo$^{3}$ \thanks{E-mail:fadulora@uis.edu.co (FDLC)} and 
        F. S. Guzm\'an$^{1}$ \thanks{E-mail:guzman@ifm.umich.mx (FSG)}   \\ 	
        $^{1}$Instituto de F\'{\i}sica y Matem\'{a}ticas, Universidad
        Michoacana de San Nicol\'as de Hidalgo.\\
        Edificio C3, Cd. Universitaria, 58040 Morelia, Michoac\'{a}n,
        M\'{e}xico. \\
        $^{2}$Instituto de Astronom\'{\i}a, Universidad Nacional Aut\'{o}noma 
        de M\'{e}xico, AP 70-264, Distrito Federal 04510, M\'{e}xico. \\
        $^{3}$Grupo de Investigaci\'on en Relatividad y Gravitaci\'on,
        Escuela de F\'isica, Universidad Industrial de Santander, \\
        A. A. 678, Bucaramanga 680002, Colombia.  
        }

\begin{document}

% --->   DATE

\date{\today}

\pagerange{\pageref{firstpage}--\pageref{lastpage}} \pubyear{2015}

\maketitle

\label{firstpage}

% -----> ABSTRACT

\begin{abstract}
We present a new code designed to solve the equations of classical ideal magnetohydrodynamics (MHD) in three dimensions, submitted to a constant gravitational field.
The purpose of the code centers on the analysis of solar phenomena within the photosphere-corona region. We present 1D and 2D standard tests to demonstrate the quality of the numerical results obtained with our code. As solar tests we present the transverse oscillations of Alfv\'enic pulses in coronal loops using a 2.5 D model, and as 3D tests we present the propagation of impulsively generated MHD-gravity waves and vortices in the solar atmosphere.
The code is based on high-resolution shock-capturing methods, uses the HLLE flux formula combined with Minmod, MC and WENO5 reconstructors.  The divergence free magnetic field constraint is controlled using the Flux Constrained Transport method.
  
\end{abstract}

% ----->   PACS


% ------------------     SECTION     ----------------------
\section{Introduction}
\label{sec:introduction}
% ------------------------------------------------------------

In solar physics one of the most important subjects is related to the dynamics of the plasma in the solar atmosphere. Specifically, phenomena which are related to the propagation of magnetohydrodynamic waves (e. g. Alfv\'en, slow, fast magnetoacoustic), and transients (e. g. vortices, jets, spicules). Recent observations based on data provided by current space missions, such as Transition  Region and Coronal Explorer (TRACE), Solar Dynamics Observatory (SDO), Solar Optical Telescope (SOT), X-ray Telescope (XRT) and Swedish Solar Telescope (SST) 
\cite{Banerjee_et_al_2007,Tomczyk_et_al_2007,De_Pontieu_et_al_2007,Okamoto_et_al_2007,Jess_et_al_2009} have enhanced the study of these phenomena in the solar atmosphere.

The type of effects in the solar atmosphere are considered when the plasma is fully ionized and therefore the ideal MHD equations represent a good model. However in this limit, these equations are highly non-linear and numerical simulations are essential, where accurate numerical algorithms play an important role. Over the recent years a number of codes have been used to understand various effects in the solar atmosphere. For instance, in \cite{Konkol&Murawski_2010,Danilko_et_al_2012} the authors used a modification of ATHENA code \cite{Stone_et_al_2008} to study vertical oscillations of solar coronal loops and magnetoacoustic waves in a corollarygravitationally stratified solar atmosphere. In \cite{Konkol_et_al_2010,Murawski&Musielak_2010,Murawski_et_al_2011,Murawski_et_al_2013,Chmielewski_et_al_2014a} the authors use a modification of the FLASH code \cite{Fryxell_et_al_2000,Lee&Deane_2009,Lee_2013} to simulate the propagation of MHD waves in two and three dimensions. Moreover in \cite{Woloszkiewicz_et_al_2014} the authors study the propagation of Alfv\'en waves in the solar atmosphere using the PLUTO code \cite{Mignone_et_al_2007}. 

Wave phenomena in the solar atmosphere have influenced the development of numerical codes dedicated specifically to study MHD wave propagation, e.g. the VAC code \cite{Toth_2000,Shelyag_et_al_2008}, which is capable to simulate the interaction of any arbitrary perturbation with a magnetohydrostatic equilibrium background. The SURYA code \cite{Fuchs_et_al_2011}, allows to study the propagation of waves in a stratified non-isothermal magnetic solar atmosphere, in particular in the chromosphere-corona interface. Numerical simulations also permit to analyze some transient effects, for example in  \cite{Murawski&Zaqarashvili_2010,Murawski_et_al_2011} using FLASH, the authors simulate the formation of spicules and macrospicules in the solar atmosphere. In addition in \cite{Pariat_et_al_2009,Pariat_et_al_2010,Pariat_et_al_2015} the authors propose a model for jet activity that is commonly observed in solar corona. Also the 3D ideal MHD equations are used  to model the magnetic reconnection producing massive and high-speed jets \cite{DeVore_1991}.

Even though all the aforementioned codes are available, well documented and can be applied to various problems, we present our new application, with convergence tests and error analyses. Our code is at the moment useful to study solar phenomena in the photosphere-corona region, however with plan to generalize it to deal with magnetic reconnection and solar jets including radiative processes.

The paper is organized as follows. In Section \ref{sec:mhd} we describe the ideal MHD equations and mention the numerical methods implemented in our code. In Section \ref{sec:tests} we show the standard tests in 1D and 2D dimensions. As a solar test in 2D we study the transverse oscillations of Alfv\'enic pulses in the solar coronal loops. We also present as a 3D test the propagation of MHD waves in a gravitationally non-isothermal solar atmosphere. Finally, in Section \ref{sec:conclusions} we mention some final comments.

% ------------------     SECTION     ----------------------
\section{MHD Equations and Numerical Methods}
\label{sec:mhd}

% ----->     Subsection     <-----
\subsection{Ideal MHD Equations}
We consider the set of the ideal MHD equations submitted to an external constant gravitational field. We choose to write down these equations in a conservative form, which is optimal for our numerical methods:

\begin{eqnarray}
&&\frac{\partial\rho}{\partial t} + \nabla\cdot(\rho{\bf v}) = 0, \label{density}\\
&&\frac{\partial(\rho{\bf v})}{\partial t} + \nabla\cdot(\rho{\bf v}\otimes{\bf v}-{\bf B}\otimes{\bf B} + {\bf I}p_{t}) =\rho{\bf g}   \label{momentum}, \\
&&\frac{\partial E}{\partial t} +\nabla\cdot((E+p_{t}){\bf v}-{\bf B}({\bf v}\cdot{\bf B})) = \rho{\bf v}\cdot{\bf g}, \label{energy} \\
&&\frac{\partial{\bf B}}{\partial t} +\nabla\cdot({\bf v}\otimes{\bf B} -{\bf B}\otimes{\bf v}) = {\bf 0}, \label{evolB} \\
&&\nabla\cdot{\bf B} = 0, \label{divergenceB} 
\end{eqnarray}

\noindent where $\rho$ is the plasma mass density, $p_{t}=p+\frac{{\bf B}^{2}}{2}$ is the total (thermal + magnetic) pressure, ${\bf v}$ represents the plasma velocity field, ${\bf B}$ is the magnetic field, ${\bf I}$ is the unit matrix and $E$ is the total energy density, that is, the sum of the internal, kinetic, and magnetic energy density

\begin{equation}
E = \frac{p}{\gamma-1} + \frac{\rho{\bf v}^{2}}{2} + \frac{{\bf B}^{2}}{2}. \label{total_energy}
\end{equation}

\noindent The internal energy term is calculated considering an equation of state of an ideal gas

\begin{equation}
p = (\gamma-1)\rho e, \label{eos}  
\end{equation}

\noindent where $e$ is the internal energy density and $\gamma$ is the adiabatic index. This equation also serves to close the system of equations (\ref{density})-(\ref{divergenceB}). In solar surface problems one considers the gravitational source term given by ${\bf g}=[0,0,-g]$ with magnitude $g=274$ m/$s^{2}$, which is the value of the constant gravitational acceleration on the solar surface. These equations are written in units such that $\mu_{0}=1$. It is also convenient to write the equation of state in terms of the temperature, that is

\begin{equation}
p =\frac{k_{B}}{m}\rho T, \label{eos_temp}  
\end{equation}

\noindent where $k_{B}$ is Boltzmann's constant, $m$ is the mean particle mass defined in terms of the mean atomic weight $\mu$ in the following way $\mu=m/m_{p}$, where $m_{p}$ is the proton mass. The above mentioned system of equations (\ref{density})-(\ref{divergenceB}) can be written  
as a first-order hyperbolic system of flux-conservative equations, which in cartesian coordinates reads

\begin{eqnarray}
\frac{\partial {\bf Q}}{\partial t} + \frac{\partial {\bf F}}{\partial x} + \frac{\partial {\bf G}}{\partial y}
+ \frac{\partial {\bf H}}{\partial z} &=& {\bf S}, \label{eq:conservative}\\
\nabla\cdot{\bf B} &=& 0, \nonumber
\end{eqnarray}

\noindent with ${\bf Q}$ a vector of conservative quantities 

\begin{equation}
\nonumber {\bf Q} = \left( \rho, \rho v_x, \rho v_y, \rho v_z, E, B_x, B_y, B_z \right)^{T}, 
\end{equation}

\noindent  and ${\bf F}$, ${\bf G}$ and ${\bf H}$ are the fluxes along each spatial direction

\begin{equation}
\nonumber {\bf F} = \left [
\begin{array}{c}
\rho v_x \\
\rho v_x^2 - B_x^2 + p_t \\
\rho v_x v_y - B_x B_y \\ 
\rho v_x v_z - B_x B_z \\
(E + p_t)v_x - B_x ({\bf B}\cdot{\bf v}) \\
0 \\
v_x B_y - v_y B_x \\
v_x B_z - v_z B_x
\end{array}
\right], 
\end{equation}

\begin{equation}
\nonumber {\bf G} = \left [
\begin{array}{c}
\rho v_y \\
\rho v_y v_x - B_y B_x \\ 
\rho v_y^2 - B_y^2 + p_t \\
\rho v_y v_z - B_y B_z \\
(E + p_t)v_y - B_y ({\bf B}\cdot{\bf v}) \\
v_y B_x - v_x B_y \\
0 \\
v_y B_z - v_z B_y
\end{array}
\right], 
\end{equation}

\begin{equation}
\nonumber {\bf H} = \left [
\begin{array}{c}
\rho v_z \\
\rho v_z v_x - B_z B_x \\ 
\rho v_z v_y - B_z B_y \\ 
\rho v_z^2 - B_z^2 + p_t \\
(E + p_t)v_z - B_z({\bf B}\cdot{\bf v}) \\
v_z B_x - v_x B_z \\
v_z B_y - v_y B_z \\
0 
\end{array}
\right].
\end{equation}

Finally the source vector ${\bf S}$ is given by

\begin{equation}
\nonumber {\bf S} = \left [
\begin{array}{c}
0 \\
0 \\
0 \\ 
-\rho g \\
-\rho v_z g\\
0 \\
0 \\
0
\end{array}
\right].
\end{equation}

\subsection{Characteristic Structure}

The numerical methods used to solve the MHD equations require the characteristic information from the Jacobian matrices associated to the system of equations (\ref{eq:conservative}) 

\begin{equation}
\nonumber A_x({\bf Q}) = \frac{\partial {\bf F}}{\partial {\bf Q}}, ~~ A_y({\bf Q}) = \frac{\partial {\bf G}}{\partial {\bf Q}}, ~~ A_z({\bf Q}) = \frac{\partial {\bf H}}{\partial {\bf Q}}.  
\end{equation}

\noindent As we can see, ${\bf G}$ and ${\bf H}$ can be obtained from an adequate index permutation of ${\bf F}$. Therefore, $A_y$ and $A_z$ have similar characteristic structure as $A_x$. The set of eigenvectors for the Jacobian matrices has been well stablished and extensively studied by several authors \cite{Roe_1986,Brio&Wu_1988,Powell_1994,Ryu&Jones_1995,Balsara_2001}. The corresponding eigenvalues for $A_x$, following the $8 \times 8$ eigen-system of Powell \cite{Powell_1994,Powell_et_al_1999} are given by

\begin{equation}
\nonumber \lambda_{1,8} = v_x \pm c_f, ~~ \lambda_{2,7} = v_x \pm c_a, ~~ \lambda_{3,6} = v_x \pm c_s, ~~ \lambda_{4,5} = v_x, 
\end{equation}

\noindent where $c_a$ is the Alfv\'en speed

\begin{equation}
\nonumber c_a = \sqrt{\frac{B_x^2}{\rho}},
\end{equation}

\noindent $c_f$ and $c_s$ are the fast and slow magnetoacoustic speeds  respectively

\begin{eqnarray}
\nonumber c_f &=& \left[\frac{1}{2}\left(a^2 + \frac{B^2}{\rho} +
\sqrt{\left(a^2 + \frac{B^2}{\rho}\right)^2 - 4a^2c_a^2}   \right) \right]^{1/2}, \\
\nonumber c_s &=& \left[\frac{1}{2}\left(a^2 + \frac{B^2}{\rho} -
\sqrt{\left(a^2 + \frac{B^2}{\rho}\right)^2 - 4a^2c_a^2}   \right) \right]^{1/2}
\end{eqnarray}

\noindent and $a = \sqrt{\frac{\gamma p}{\rho}}$ is the sound speed. Finally, similar expressions are found for $A_y$ and $A_z$.

%\begin{equation}
%\nonumber a = \sqrt{\frac{\gamma p}{\rho}}.
%\end{equation}

% ----->     Subsection     <-----
\subsection{Numerical Methods}

We solve numerically the magnetized Euler equations given by the system (\ref{density})-(\ref{divergenceB}) on a single uniform cell centered grid,  using the method of lines with a third order total variation diminishing Runge-Kutta time integrator (RK3) \cite{Shu&Osher_1989}. In order to use the method of lines, the right hand sides of MHD equations are discretized using a finite volume approximation with High Resolution Shock Capturing methods. For this, we first reconstruct the variables at cell interfaces using the Minmod and MC linear piecewise second-order reconstructors, and the fifth-order weighted Essentially Non Oscillatory (WENO5) method \cite{Titarev&Toro_2004,Harten_1997,Radice&Rezzolla_2012}, that uses a polynomial interpolation with a fifth order precision for smooth distributions and third order at strong shocks. Even though the accuracy of the RK3 dominates over a high order accuracy reconstructor during the time integration, we include WENO5 because it introduces a small amount of dissipation.

On the other hand, the numerical fluxes are built with the Harten-Lax-van-Leer-Einfeldt (HLLE) approximate Riemann solver formula, using the left and right states of the primitive and conservative reconstructed variables \cite{Harten_et_al_1983,Einfeldt_1988}. The HLLE approximate solver uses a two-wave approximation to compute the fluxes across the discontinuity at the cell interfaces, including those of the magnetic and hydrodynamic waves. Such waves move  with the highest and lowest  velocities that correspond to eigenvalues of the Jacobian matrix. These numerical methods were implemented successfully in relativistic MHD, where we have studied different phenomena with fixed and dynamic background space-times, including the presence of black holes \cite{Lora-Clavijo2014,Lora-Clavijo2013,Lora-Clavijo2013b,Lora-Clavijo2013c,Cruz-Osorio2012,Lora-Cruz2015}.

Additionally, in order to preserve the constraint (\ref{divergenceB}) and avoid the violation of Maxwell's equations (i.e. the development of unphysical results like the presence of magnetic charges) we use the flux constraint transport (flux-CT) algorithm \cite{Evans&Hawley_1988,Balsara_2001}. The flux-CT method is adapted to the conservative scheme computing the fluxes at cell vertex as an average of the fluxes at intercell boundaries and resulting in a cell-faced evolution equation of the components of the magnetic field. This scheme maintains the divergence of the magnetic field computed at cell corners to round-off accuracy along the numerical evolution. The details of the numerical implementation can be found in \cite{cafe} and other divergence constraint control methods are summarized in \cite{Toth_2000}.

% ------------------------------------------------------------

% ------------------     SECTION     ----------------------
\section{Numerical Tests}
\label{sec:tests}

We include a set of 1D and 2D tests that involve the evolution with the gravitational field switched off. The solar tests include 2.5D and 3D tests involving the propagation of MHD waves in the solar atmosphere submitted to a constant gravitational field along the vertical direction. 
% ------------------------------------------------------------

% ------------------------------------------------------------
% ----->     Subsection     <-----
\subsection{1D tests}

In order to illustrate how our implementation handles the evolution of a magnetized gas, in this subsection we present the standard shock tube tests in one direction. The 1D tests are Riemann problems under various initial conditions and we compare the numerical results with the exact solution calculated in \cite{Takahashi&Yamada_2013,Takahashi&Yamada_2014}.

We calculate the numerical solution of these tests using the HLLE formula and the various limiters mentioned, in the domain $[-0.5,0.5]$ with $N = 1600$ identical cells, a constant Courant factor $CFL=0.25$ for the first test, and $0.05$ for the others. The reason to choose this value is that in order to compare among reconstructors used here, we use a value that can be handled by all of them. The initial discontinuity in this case is located at $x=0$. In order to estimate the error between the numerical solution and the exact solution, we calculate the numerical solution using four spatial resolutions $\Delta x_1=1/200$, $\Delta x_2=1/400$, $\Delta x_3=1/800$ and $\Delta x_4=1/1600$. For these tests we are using the 3D code with four cells along the transverse directions. The various parameters of 1D tests are summarized in Table \ref{tab:MHD_tests} and we practiced this test for the three limiters mentioned.

\begin{table}
\centering
\resizebox{9cm}{!}{
\begin{tabular}{c|c|c|c|c|c|c|c|c|c}\hline \hline
${\bf Test ~ type}$ & $\gamma$ & $\rho$ & $p$ & $v_x$ & $v_y$ & $v_z$ & $B_x$ & $B_y$ & $B_z$ \\ \hline \hline
${\bf Brio-Wu}$ &  &  &  &  &  & &  &  &  \\
Left state  & 5/3 & 1.0 & 1.0 & 0.0 & 0.0  & 0.0 & 0.75  & 1.0  & 0.0   \\ 
Right state &  & 1.0 & 1.0 & 0.0  & 0.0  & 0.0 & 0.75 & -1.0  & 0.0   \\ \hline \hline
${\bf Ryu\&Jones ~ 1b}$ &  &  &  &  &  &  &  &  &   \\
Left state  & 5/3 & 1.0 & 1.0 & 0.0 & 0.0  & 0.0 & $3/\sqrt{4\pi}$ & $5/\sqrt{4\pi}$  & 0.0   \\ 
Right state &  & 0.1 & 10.0 & 0.0  & 0.0  & 0.0 & $3/\sqrt{4\pi}$ & $2/\sqrt{4\pi}$ & 0.0   \\ \hline \hline
${\bf Ryu\&Jones ~ 2b}$ &  &  &  &  &  & &  &  &   \\
Left state  & 5/3 & 1.0 & 1.0 & 0.0 & 0.0  & 0.0 & $3/\sqrt{4\pi}$ & $6/\sqrt{4\pi}$ & 0.0 \\ 
Right state &  & 0.1 & 10.0 & 0.0  & 2.0  & 1.0 & $3/\sqrt{4\pi}$ & $1/\sqrt{4\pi}$  & 0.0    \\  \hline \hline
${\bf Ryu\&Jones ~ 3b}$ &  &  &  &  &  & &  &  &   \\
Left state  & 5/3 & 1.0 & 1.0 & -1.0 & 0.0  & 0.0 & 0.0 & 1.0 & 0.0  \\ 
Right state &  & 1.0 & 1.0 & 1.0  & 0.0  & 0.0 & 0.0  & 1.0  & 0.0   \\ \hline \hline
\end{tabular}
}
\caption{\label{tab:MHD_tests} Parameters for the various MHD 1D Riemann problems.} 
\end{table}

% ----->     Brio-Wu
\subsubsection{Test 1: Brio-Wu Shock Tube Problem}  

\begin{figure}
\begin{center}
\includegraphics[width=6.0cm]{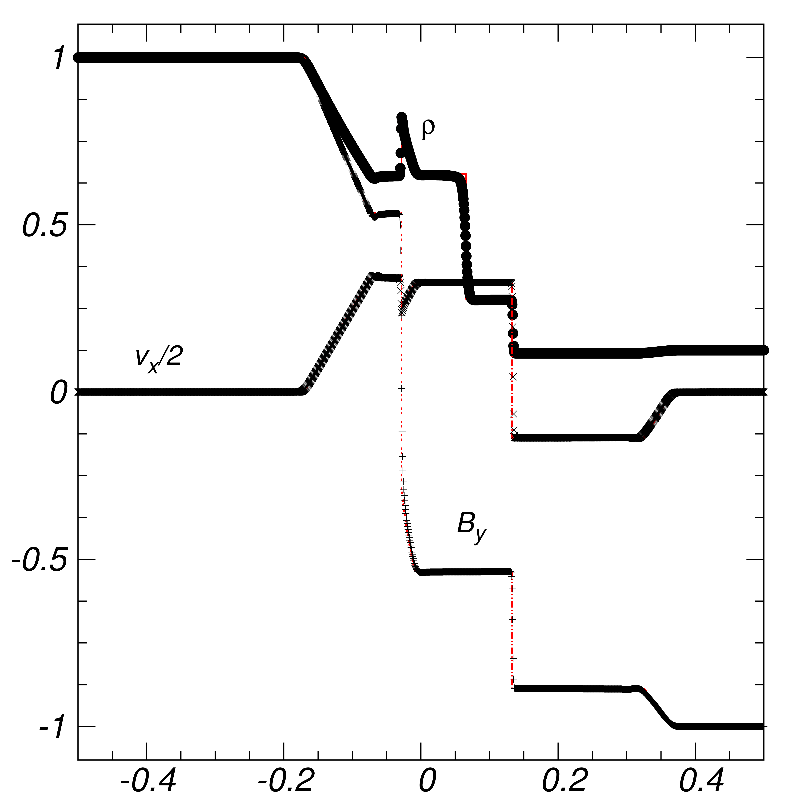}
\caption{\label{fig:Brio-Wu} Test 1: Brio-Wu shock tube problem at $t=0.1$. We show the numerical solution for $\rho$, $v_{x}$ and $B_{y}$ (points) and the exact solution (line). In this case we use the Minmod limiter.}
\end{center}
\end{figure}

The Brio-Wu MHD shock tube problem \cite{Brio&Wu_1988} tests the capability of a code to handle waves of various types, shocks, rarefactions, contact discontinuities and the compound structures of MHD. A snapshot of the wave structure at $t=0.1$ of the solution is shown in Fig. \ref{fig:Brio-Wu}. The left-moving fast rarefaction wave from $x\approx-0.2$ to $x\approx-0.1$ and the slow compound wave at $x\approx-0.03$ are well captured. The contact discontinuity  is captured at $x\approx0.05$. Finally the slow shock is captured at $x\approx0.15$ and fast rarefaction from $x\approx0.32$ to $x\approx0.37$. 

% ----->     Ryu-Jones 1b
\subsubsection{Test 2: Ryu \& Jones 1b}

\begin{figure}
\begin{center}
\includegraphics[width=6.0cm]{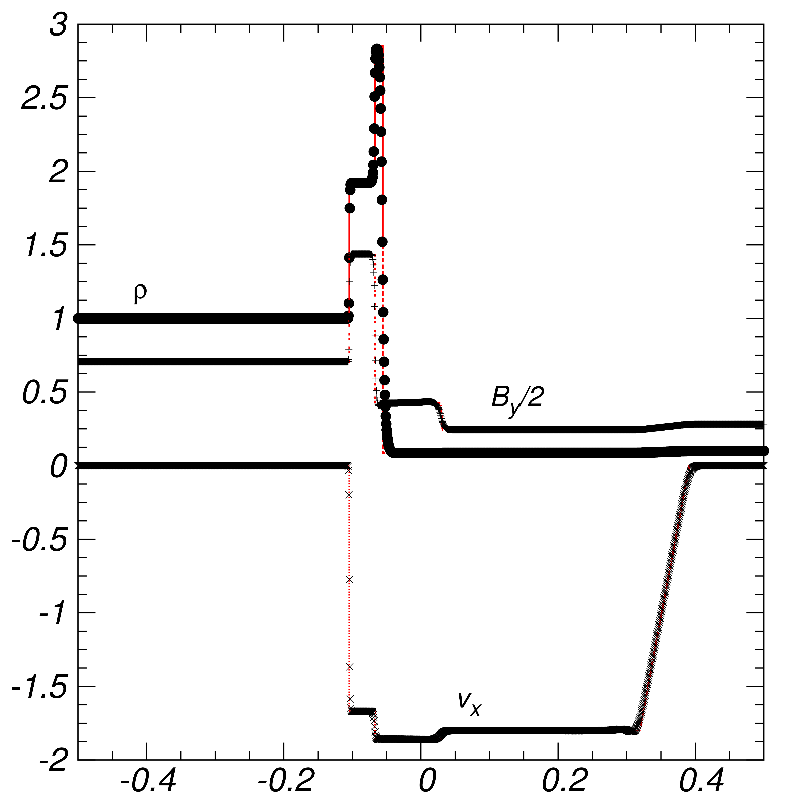}
\caption{\label{fig:Ryu-Jones_1b} Test 2: Ryu \& Jones 1b shock tube problem at $t=0.03$. We show the numerical solution for $\rho$, $v_{x}$ and $B_y$ (points) compared with the exact solution (line). Again we use the Minmod limiter.}
\end{center}
\end{figure}

This test was introduced by \cite{Ryu&Jones_1995}. It is important to check on the ability of the numerical scheme to handle a fast shock, a rarefaction shock, a slow shock, a slow rarefaction, and a contact discontinuity. A snapshot of the evolution is displayed in Fig. \ref{fig:Ryu-Jones_1b}. Our results show that fast shock at $x\approx-0.11$ and the slow shock at $x\approx-0.075$ are well captured. In addition the contact discontinuity appears at $x\approx-0.061$ and the slow rarefaction is located at $x\approx-0.040$. Finally the fast rarefaction is captured at $x\approx0.34$.

% ----->     Ryu-Jones 3b
\subsubsection{Test 3: Ryu \& Jones 3b}

\begin{figure}
\begin{center}
\includegraphics[width=6.0cm]{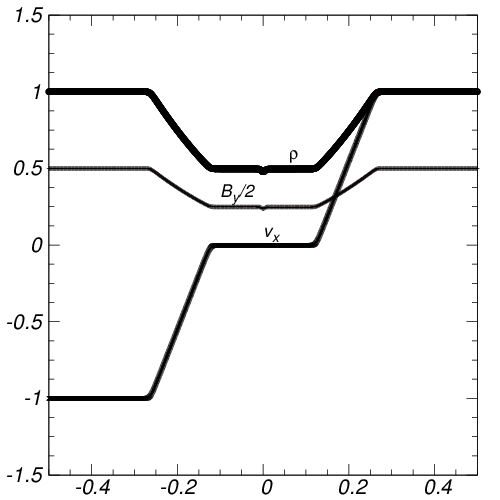}
\caption{\label{fig:Ryu-Jones_3b} Test 3: Ryu \& Jones 3b shock tube problem at $t=0.1$. We show the numerical solution for $\rho$, $v_x$ and magnetic field $B_{y}$ (points) compared with the exact solution (line).}
\end{center}
\end{figure}

This is a test containing magnetosonic rarefactions waves. The numerical solution shows only two strong and identical magnetosonic rarefractions. Our code was able to capture this structure and the results are shown in Fig. \ref{fig:Ryu-Jones_3b}.   

% -----> Errors

\subsubsection{Error estimates for the 1D MHD tests}

In order to compare the numerical solutions using different limiters implemented in our code, we have calculated the error for each 1D Riemann test described above, and the results are summarized in Table \ref{tab:mhd_error}. In all the tests, with the three reconstructors used, our runs lie within the consistency regime, that is, the error decreases when resolution is increased. In order to go further we check for the order of convergence. For this we calculate the $L_1$ norm of the error  as compared with the exact solution. Since we always consider resolution factors of two, the order of convergence is given by $\log(error_i/error_{i-1})/\log(2)$, where $i$ is the error calculated with the resolution $\Delta x_i$. Thus we carried out the tests with four spatial resolutions. For all the tests we obtained near first order convergence as expected, considering all the initial data start with discontinuities and in Table \ref{tab:mhd_error} we show how the different reconstructors perform in the different tests.

\begin{table*}
\begin{center}
\begin{tabular}{c||c|c|c|c||c|c|c|c} \hline \hline
 ${\it Resolution}$ & MM & MC & WENO5 & MM & MC & WENO5 \\ \hline \hline
\multicolumn{1}{|c}{} & \multicolumn{2}{c|}{Error} & & \multicolumn{2}{|c|}{Order of convergence} \\ \hline \hline
\multicolumn{1}{|c}{} & \multicolumn{5}{c|}{${\bf Brio-Wu}$} \\ \hline \hline
$\Delta x_1$   & 0.11e-1 & 0.11e-1   &  0.82e-2   & ....  & ....   & .... \\ \hline \hline
$\Delta x_2$   & 0.66e-2 & 0.66e-2   &  0.45e-2   & 0.74  &  0.74  & 0.86  \\ \hline \hline
$\Delta x_3$   & 0.41e-2 & 0.41e-2   &  0.28e-2   & 0.69  &  0.69  & 0.68   \\ \hline \hline
$\Delta x_4$   & 0.25e-2 & 0.25e-2   &  0.19e-2   & 0.71  &  0.71  & 0.55    \\ \hline \hline
\multicolumn{1}{|c}{} & \multicolumn{5}{c|}{${\bf Ryu \& Jones ~ 1b}$} & \\ \hline \hline
$\Delta x_1$   & 0.32e-1 & 0.27e-1   & 0.24e-1    &  .... & ....  & .... \\ \hline \hline
$\Delta x_2$   & 0.22e-1 & 0.15e-1   & 0.13e-1    &  0.71 & 0.84  & 0.88  \\ \hline \hline
$\Delta x_3$   & 0.14e-1 & 0.80e-2   & 0.74e-2    &  0.65 & 0.90  & 0.81  \\ \hline \hline
$\Delta x_4$   & 0.86e-2 & 0.50e-2   & 0.45e-2    &  0.70 & 0.67  & 0.71   \\ \hline \hline
\multicolumn{1}{|c}{} & \multicolumn{5}{c|}{${\bf Ryu \& Jones ~ 3b}$} \\ \hline \hline
$\Delta x_1$   & 0.71e-2 & 0.46e-2   & 0.43e-2    & ....  & .... & .... \\ \hline \hline
$\Delta x_2$   & 0.37e-2 & 0.23e-2   & 0.22e-2    & 0.94  & 1.0  & 0.96  \\ \hline \hline
$\Delta x_3$   & 0.19e-2 & 0.12e-2   & 0.12e-2    & 0.96  & 0.93 & 0.87  \\ \hline \hline
$\Delta x_4$   & 0.10e-2 & 0.67e-3   & 0.79e-3    & 0.92  & 0.84 & 0.60   \\ \hline \hline
\end{tabular}
\end{center}
\caption{\label{tab:mhd_error} $L_1$ norm of the error in density calculated with four resolutions and three numerical reconstructors. We also show the order of convergence between the different pairs of resolutions.}
\end{table*}
%\end{center}
%\end{widetext}

% ------------------------------------------------------------
% ----->     Subsection 2D TESTS    <-----
\subsection{2D tests}

\subsubsection{Orszag-Tang Vortex} 

The Orszag Tang vortex system was first studied by \cite{Orszag&Tang_1998}. This is a well known problem for testing the transition to a supersonic 2D MHD turbulence, and to measure the robustness of the code at handling the formation of MHD shocks. The important dynamics developed during the evolution is a test for the controller of the magnetic field divergence free constraint $\nabla\cdot B=0$.

The initial data are ${\bf v}= v_{0}(-\sin(2\pi y),\sin(2\pi x),0)$, ${\bf B}= B_{0}(-\sin(2\pi y),\sin(4\pi x),0)$, $\rho_0 =25/(36\pi) $ and $p_0 = 5/(12\pi)$, where $v_{0}=1$ and $B_{0}=1/\sqrt{4\pi})$ with $\gamma=5/3$, 
are defined in such a way that the system develops a vortex. The initial data for $\rho$ and $p$ are calculated using averaged $\beta$ and the Mach number $M$ \cite{Dai&Woodward_1998}, which are defined as 

\begin{eqnarray}
\beta &=& \frac{p_0}{<{\bf B}^{2}>/8\pi}=8\pi p_0, \nonumber \\
 M &=&\frac{{<\bf u^{2}>}}{c_s^{2}}=\frac{\rho_0}{\gamma p_0}. \nonumber
\end{eqnarray} 

In this case $\beta=10/3$ and $M=1$. We carry out the evolution on the domain $[0,1]\times[0,1]$ that we cover with $512 \times 512$ cells, and impose periodic boundary conditions.

\begin{figure}
\begin{center}
\includegraphics[width=6.0cm]{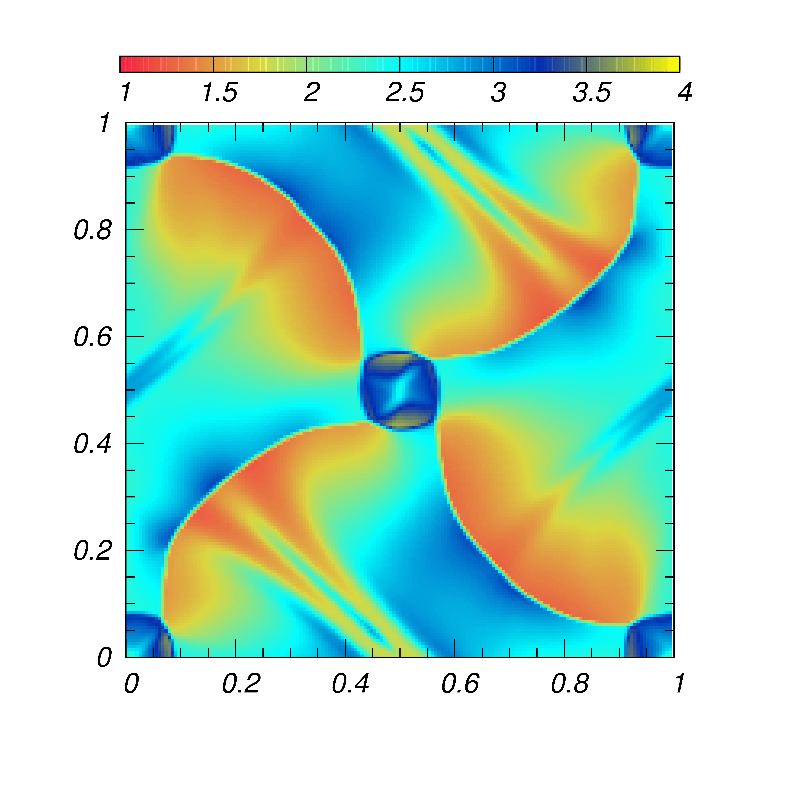}
\includegraphics[width=6.0cm]{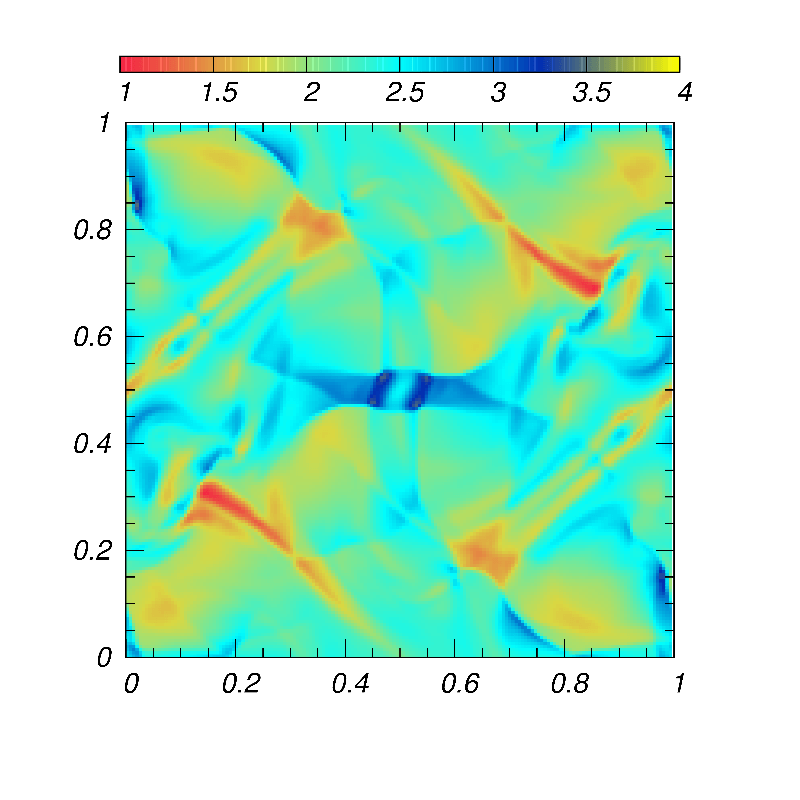}
\includegraphics[width=6.0cm]{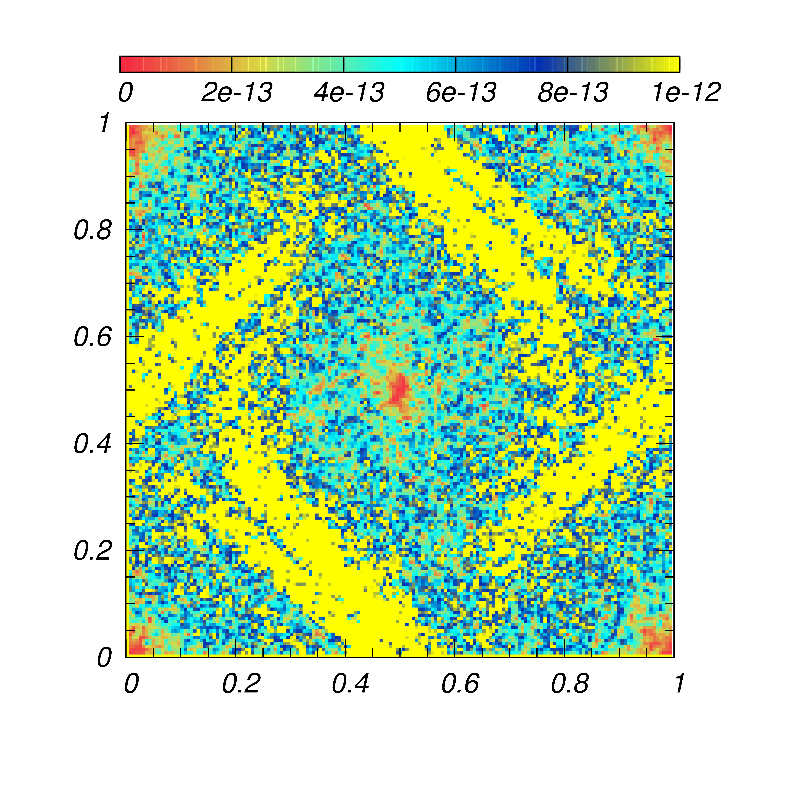}
\caption{\label{fig:rho_OTVt5} Density for the Orszag-Tang vortex at times $t=0.5$ (top) and $t=1.0$ (middle). We show $|\nabla\cdot B|$ at $t=1.0$ in the $xy$ plane (bottom).}
\end{center}
\end{figure}

\begin{figure}
\begin{center}
\includegraphics[width=6.0cm]{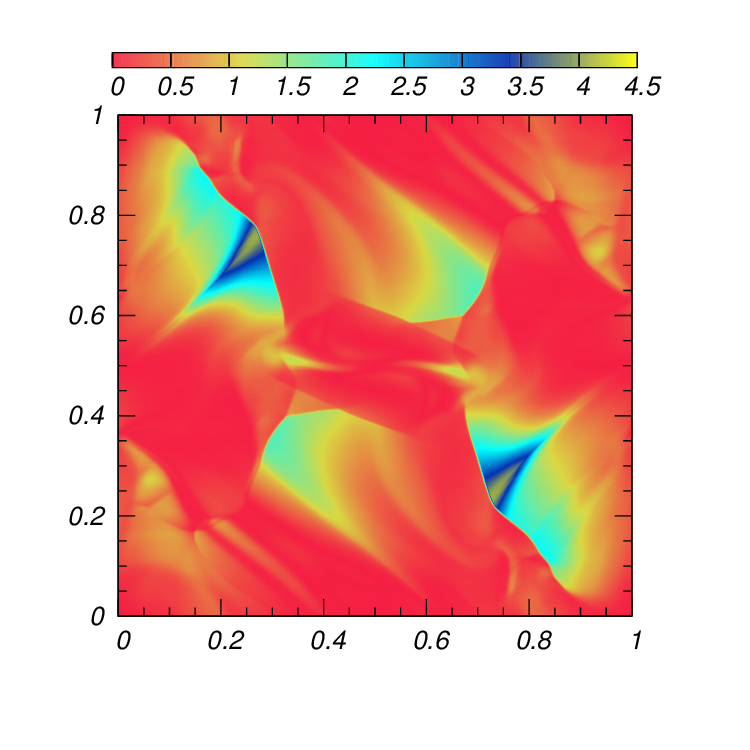}
\includegraphics[width=6.0cm]{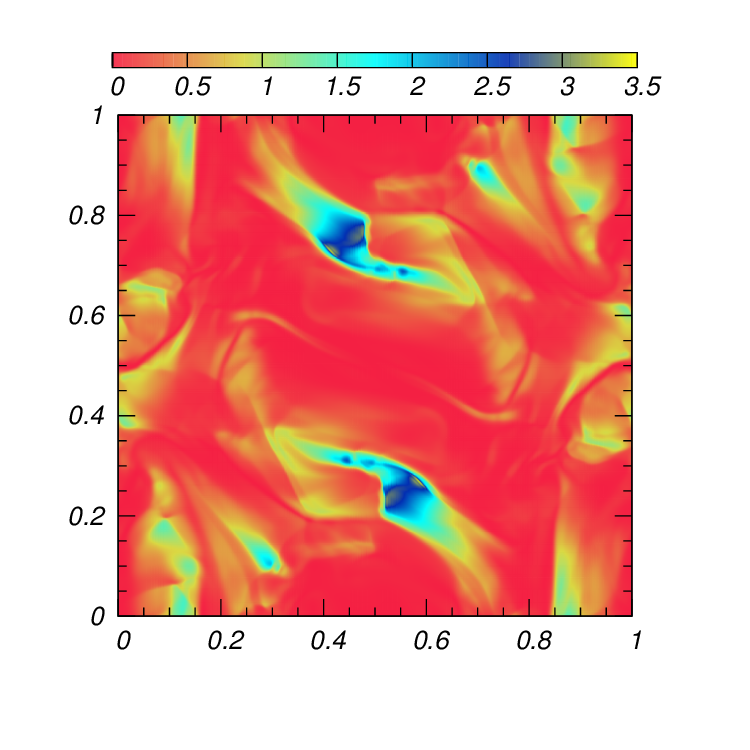}
\caption{\label{fig:Nmach_OTV} Mach number $M$ for the Orszag-Tang vortex at times $t=0.5$ (top) and $t=1.0$ (bottom) in the $xy$ plane. The supersonic regions are located in the dark colored zones.}
\end{center}
\end{figure}

\begin{figure}
\begin{center}
\includegraphics[width=9.0cm]{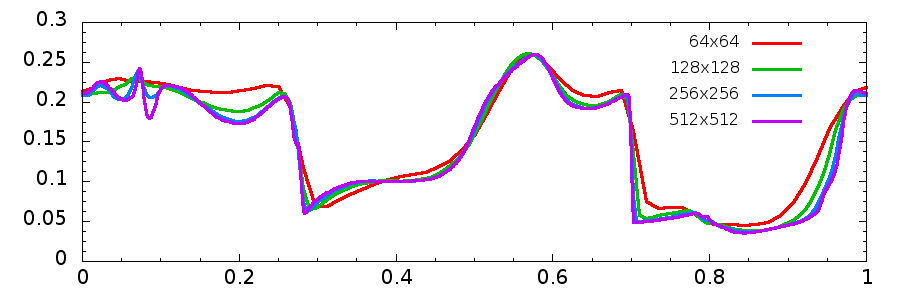}
\caption{\label{fig:p_cut_y03125} Horizontal slices of $p$ along $y=0.3125$ for the Orszag-Tang vortex using four spatial resolutions using Minmod. This plot shows the consistency of the solution.}
\end{center}
\end{figure}

The evolution of the density shows the development of turbulence and the interaction of the shock within the domain as shown in the snapshots taken at times $t=0.5$ and $t=1.0$ in Fig. \ref{fig:rho_OTVt5}, which can be compared with the results from ATHENA \cite{Stone_et_al_2008}. Our code maintains the constraint on values $|\nabla\cdot B|\approx 10^{-13}$, which is within the round-off error of the machine accuracy. In order to show how the fluid makes the transition to a supersonic turbulence, we calculate the Mach number defined as $M=|v|^{2}/c_s^{2}$, where $|v|^{2}=v_x^{2}+v_y^{2}$, where $c_s^{2}=\frac{\gamma p}{\rho}$ is the speed of sound. Snapshots taken at times $t=0.5$ and $t=1$ of $M$ are shown in Fig. \ref{fig:Nmach_OTV}, where we can see that Mach number increases in regions where the density decreases. More quantitative results are shown in Fig. \ref{fig:p_cut_y03125}, where the pressure with various resolutions is measured on the line $y=0.3125$ at time $t=0.5$. The results are shown for the Minmod reconstructor and are in agreement with those in Fig. 11 of \cite{Londrillo&Del_Zanna_2000}. Moreover, the results can be compared with those using FLASH \cite{Fryxell_et_al_2000,Lee&Deane_2009,Lee_2013}, ATHENA \cite{Stone_et_al_2008} and PLUTO \cite{Mignone_et_al_2007}. %Unlike the mentioned papers we show here the violation of the constarint. 

%% ----------------------- Circularly polarized Alfvén waves ------------------- %%

\subsubsection{The circularly polarized Alfv\'en Wave Test}

This test problem was presented in \cite{Toth_2000}. The circularly polarized wave is an exact nonlinear solution to the MHD equations, which allows one to test a code in the nonlinear regime. This test is useful to measure the ability of the numerical scheme to describe the propagation of Alfv\'en waves. The initial data of this problem are $\rho = 1$, $p = 0.1$, $v_{\perp} = 0.1\sin(2\pi x_{\parallel})$, $v_{\parallel} = 0$, $B_{\perp} = 0.1\sin(2\pi x_{\parallel})$, 
$B_{\parallel} = 1$, $v_{z} = 0.1\cos(2\pi x_{\parallel})$, $B_{z} = 0.1\cos(2\pi x_{\parallel})$, where $x_{\parallel}=(x\cos\alpha + y\sin\alpha)$ and $\alpha=30^{\circ}$ is the angle at which the wave propagates with respect to the grid. Here $v_{\perp}$ and $B_{\perp}$ are the components of velocity and magnetic field perpendicular to the wavevector. The Alfv\'en speed is $|v_{A}|=B_{\parallel}/\rho=1$, which implies that at time $t=1$ the flow is expected to return to its initial state. 

\begin{figure}
\begin{center}
\includegraphics[width=5.0cm]{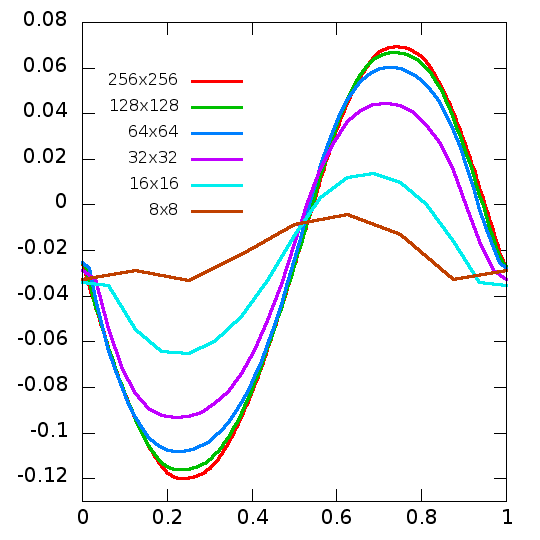}
\includegraphics[width=5.0cm]{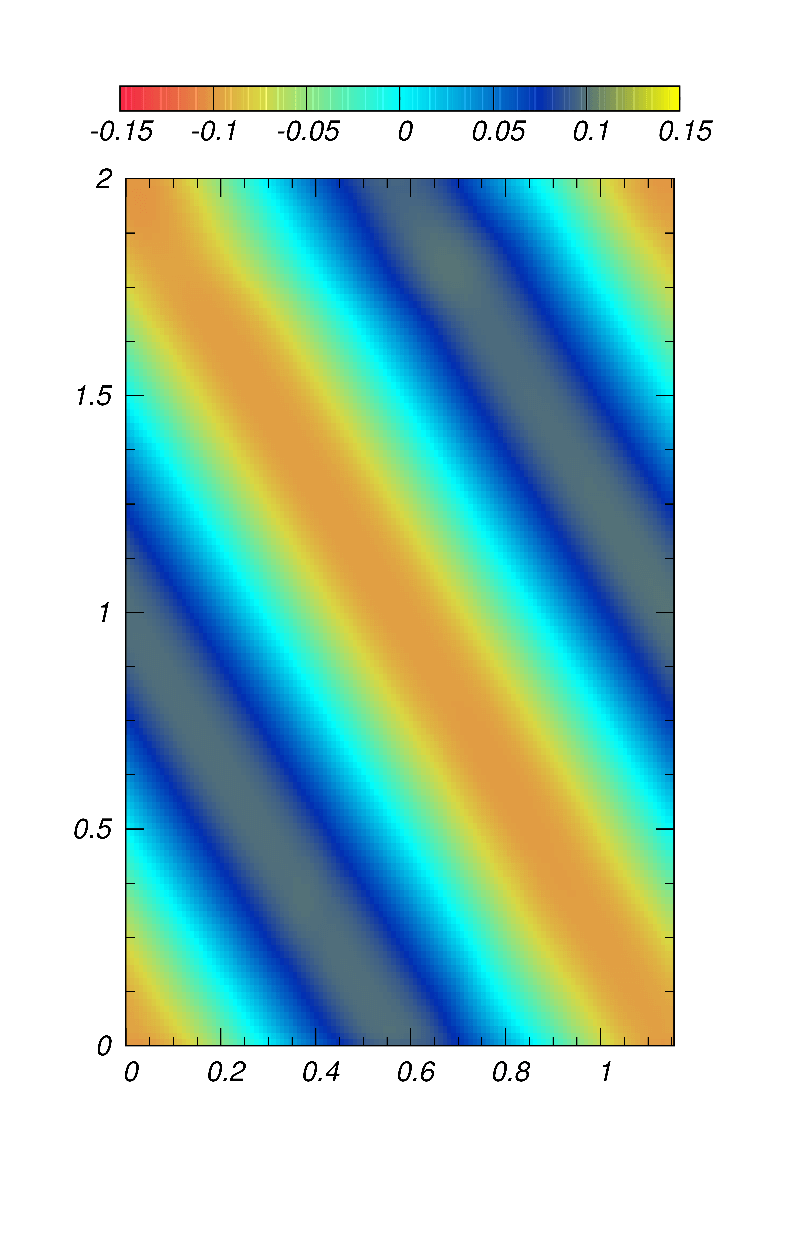}
\includegraphics[width=5.0cm]{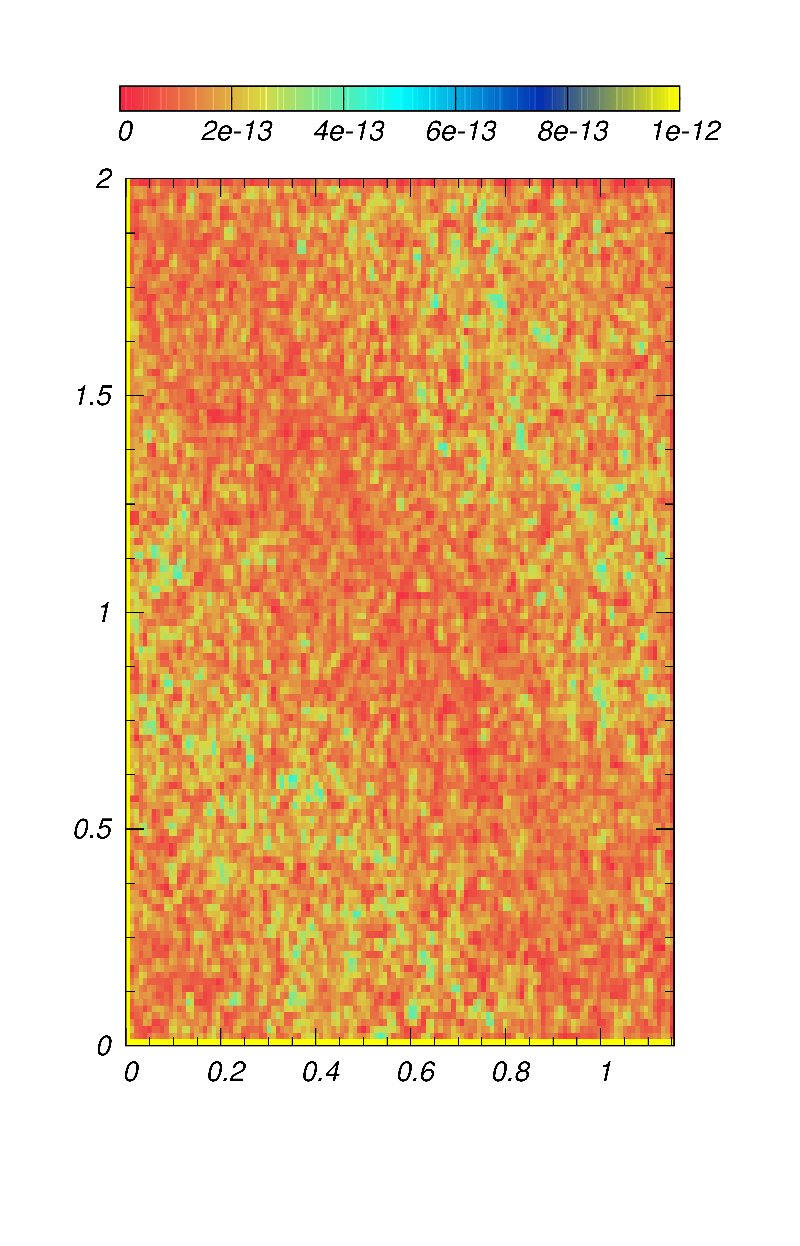}
\caption{\label{fig:Bperp_CPAW} Circularly polarized Alfv\'en wave test. We show the perpendicular component of magnetic field $B_{\perp}$ as a function of $x_{\parallel}$ at time $t=5$, using six resolutions (top), $B_z$ at time $t=5$ (middle) and $|\nabla\cdot B|$ at $t=5$ (bottom) in the $xy$ plane.}
\end{center}
\end{figure}

The domain $[0,1/\cos\alpha]\times[0,1/\sin\alpha]$ is covered with $256 \times 256$ cells. In order to see that our code is capable to describe the propagation of circularly polarized Alfv\'en waves along the diagonal grid, in Fig. \ref{fig:Bperp_CPAW} (top) we show a snapshot of $B_{\perp}$ as function of $v_{\parallel}$ taken at time $t=5$ for 6 different resolutions in the case of a travelling wave ($v_{\parallel}=0$) (top), in this case the diffusion of the wave is less for a higher resolution as expected. We also show a snapshot of $B_z$ taken at time $t=5$, where one can notice that the wavefronts remain planar despite the propagation occurs along a non-preferred grid direction. Again the code maintains $|\nabla\cdot B|\approx10^{-13}$ as shown in Fig. \ref{fig:Bperp_CPAW}. In this test we use Minmod and the results can be compared with those obtained with ATHENA \cite{Stone_et_al_2008}.

%% ----------------------- Current Sheet ------------------- %%

\subsubsection{Current Sheet Test problem}

This two dimensional problem was presented in \cite{Hawley&Stone_1995} and is important because it shows the ability of the code to deal with magnetic reconnection driven solely by numerical resistivity. Additionally, the fact that magnetic reconnection produces high pressure gradients this test also serves to monitor the behavior of the code when dealing with these sharp gradients.

\begin{figure*}
\begin{center}
\includegraphics[width=6.0cm]{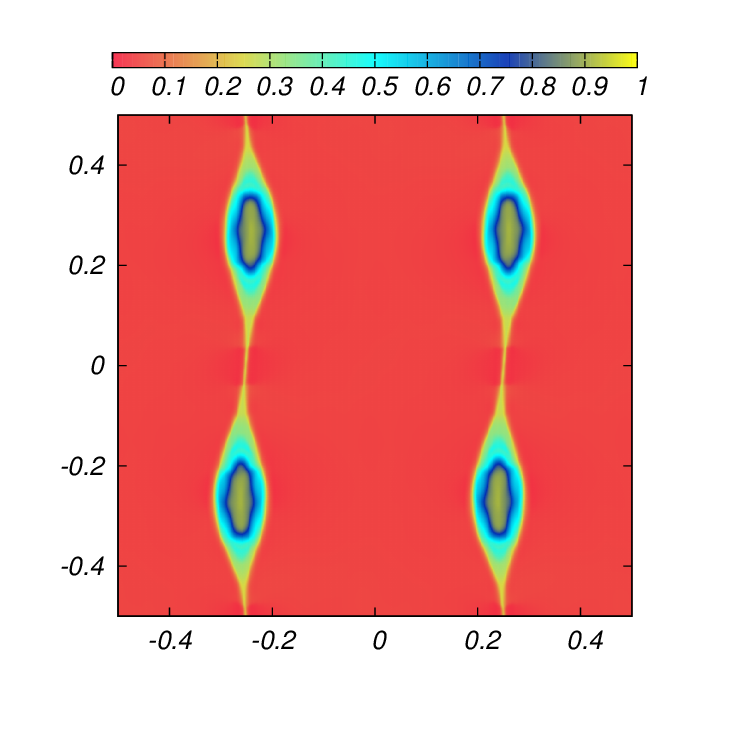}
\includegraphics[width=6.0cm]{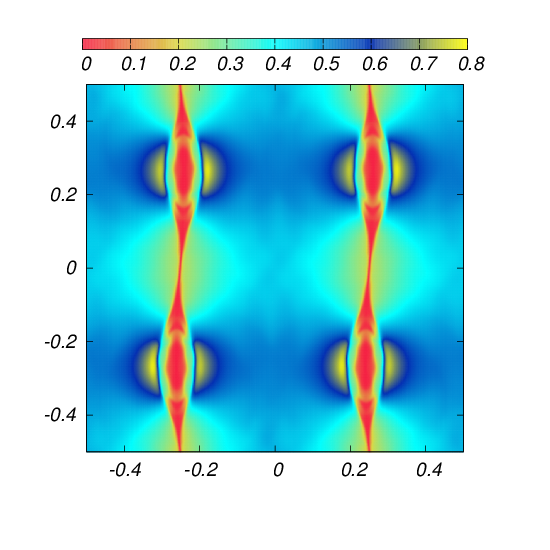}
\includegraphics[width=6.2cm]{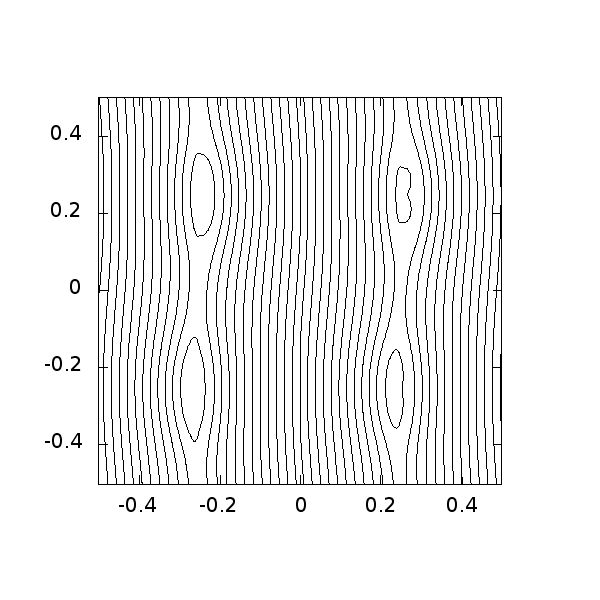}
\includegraphics[width=6.0cm]{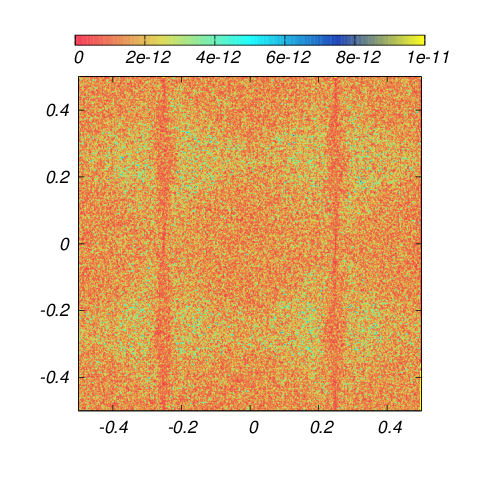}
\caption{\label{fig:current_sheet} Current sheet test. We show gas pressure $p$, magnetic pressure $B^{2}/2$, magnetic field lines and $|\nabla\cdot B|$ at time $t=1.5$ in the $xy$ plane.}
\end{center}
\end{figure*}

The computational domain $[-0.5,0.5]\times[-0.5,0.5]$ is covered with $600\times 600$ cells. We use the Minmod  reconstructor and impose periodic boundary conditions in all sides. We initialize the two current sheets as follows:

\begin{equation}
B_{y} = \left\{
\begin{array}{l@{\quad}l}
B_{0}\quad for & |x|> 0.25 \\
-B_{0} & otherwise,  
\end{array}
\right. \nonumber
\end{equation}

\noindent where $B_{0}=1$. The other magnetic field components $B_{x}$, $B_{z}$ are set to zero. The velocity is ${\bf v}=(u_0\sin(2\pi y),0,0)$ with $u_{0}=0.1$. The density $\rho=1$ and the gas pressure $p=\beta/2$, in this case we choose $\beta=0.1$. The problem is initialized with a magnetic field along the $y$ direction, that switches signs twice in the domain. This system is then perturbed with a sinusoidal velocity function, which initially creates nonlinear polarized Alfv\'en waves. These waves quickly turn into magnetosonic waves. Eventually, numerical reconnection occurs between the field lines as shown in Fig. \ref{fig:current_sheet} (bottom-left), which causes magnetic energy to be turned into thermal energy. This process can be seen in the plots of Fig. \ref{fig:current_sheet}, which show regions of high gas pressure and low magnetic pressure, these regions appear in the form of bubbles which eventually merge with each other.  Moreover, even though reconnection takes place, the constraint is kept under control $|\nabla\cdot B|\approx10^{-12}$. The results can be compared with those using FLASH \cite{Fryxell_et_al_2000,Lee&Deane_2009,Lee_2013} and ATHENA \cite{Stone_et_al_2008}.

% ----->     Subsection     <-----
\subsection{Solar Tests}

In order to show that our code can handle solar physics problems in the gravitationally stratified solar atmosphere, we study the transverse oscillations of Alfv\'enic pulses in solar coronal loops using a 2.5 D model, and in the case of a 3D configuration, we study the propagation of impulsively generated magnetohydrodynamic-gravity waves and vortices in the gravitationally stratified non-isothermal solar atmosphere.

%-----> Transverse oscillations in solar coronal loops ---------------------------
\subsubsection{Transverse oscillations in solar coronal loops}

Following the idea developed in \cite{Del_Zanna_et_al_2005}, we study impulsively generated Alfv\'en waves propagating in a magnetic arcade, taking into account a stratified solar atmosphere with height from the photosphere-corona region in order to study transverse oscillations in coronal loops. These oscillations are considered due to flare events generally interpreted as standing kink oscillations.

In this problem, the model of the coronal arcade is obtained from a solution of the static 2D MHD equations \cite{Priest_1982}, and is given by:

\begin{eqnarray}
B_x &=& B_0\cos(kx)\exp(-kz), \nonumber\\
B_z &=& -B_0\sin(kx)\exp(-kz), \label{coronal_arcade}
\end{eqnarray}

\noindent where $B_0=40$ G is the photospheric field magnitude at the footpoints $x= \pm L/2$ and $k=\pi/L$. We choose in our set up to use $L=50$Mm. This is a force and current free magnetic field, which implies that the pressure gradient must simply balance gravity in the vertical direction, that is, according to (\ref{momentum})

\begin{equation}
-\nabla p_{e} +\rho_{e}{\bf g} = {\bf 0}. \label{equilibrium_pressure}
\end{equation}

Using the equation of state for a fully ionized hydrogen plasma $p=2k_B\rho T/m_p$, we obtain the thermal pressure:

\begin{equation}
p(z) = p(z_0)\exp\left(-\frac{m_p g}{2 k_B}\int_{z_0}^{z}\frac{dz^{\prime}}{T(z^{\prime})}\right), \label{thermal_pressure}
\end{equation}

\noindent where $p(z_0)$ is the pressure at photospheric level. Then the mass density is calculated using the equation of state (\ref{eos_temp}) and gives $\rho(z)=\frac{m_p p(z)}{2k_B T(z)}$. In this problem the temperature model for the solar atmosphere is given by a smoothed step function 

\begin{equation}
T(z) = \frac{1}{2}T_{cor}\left(1+dtc+(1-dtc)\tanh\left(\frac{z-z_t}{z_w}\right)\right), \label{Smoothed_temperature}
\end{equation} 

\noindent where $dtc=T_{phot}/T_{cor}$ and $T_{phot}=6000$ K is the temperature at the photosphere, $T_{cor}=1.2\times10^{6}$ K is the higher coronal temperature. These two regions are separated at $z_t=2$ Mm by a transition region of width $z_w=0.2$ Mm. With this profile we integrate  (\ref{thermal_pressure}) to obtain the pressure and therefore the density. 

Following \cite{Del_Zanna_et_al_2005}, in order to simulate impulsively generated Alfv\'en waves, we add a purely transversal velocity pulse $v_y$ at initial time $t=0$, located at $(x_0,z_0)$:

\begin{equation}
v_y = \frac{A v_0}{1+(r/r_0)^4} \label{transverse_pulse},
\end{equation}  

\noindent where $r=\sqrt{(x-x_0)^2+(z-z_0)^2}$ with $x_0=0$ Mm and $z_0=50$ Mm, $A=0.1$ is the normalized amplitude, $v_0=1$ Mm/s is the typical Alfv\'en speed in the corona and $r_0=1$ Mm. Therefore the initial conditions for this problem are given by : $p=p(z)$, $\rho=\rho(z)$, the velocity field ${\bf v}=(0,v_y,0)$, and ${\bf B}=(B_x,0,B_z)$. 

We solve the equations (\ref{density})-(\ref{divergenceB}) using the Minmod slope limiter, a Courant factor $CFL$=0.2 and $\gamma=5/3$. We set the simulation domain to [-25,25]$\times$[0,1]$\times$[0,50] Mm$^{3}$, covered with 400$\times$4$\times$400 cells, and we impose open boundary conditions in all faces.  

\begin{figure*}
\begin{center}
\includegraphics[width=6.0cm]{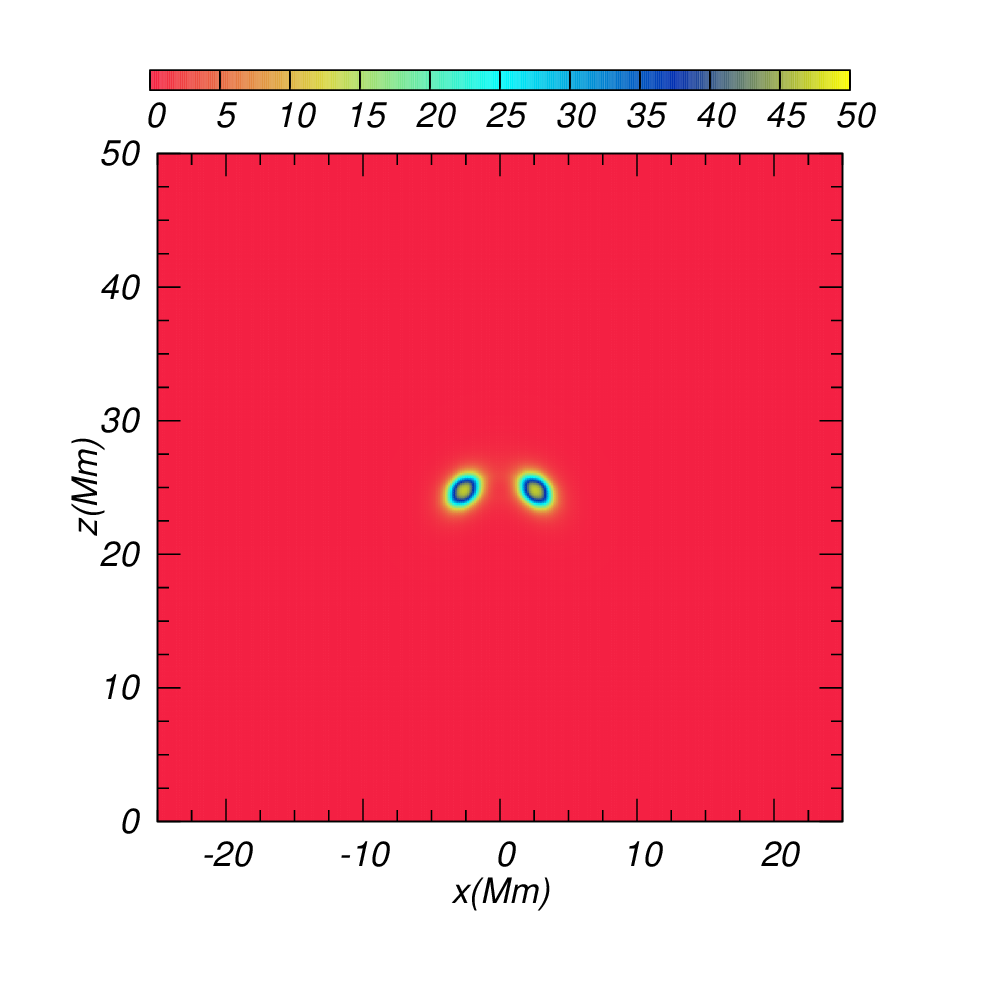}
\includegraphics[width=6.0cm]{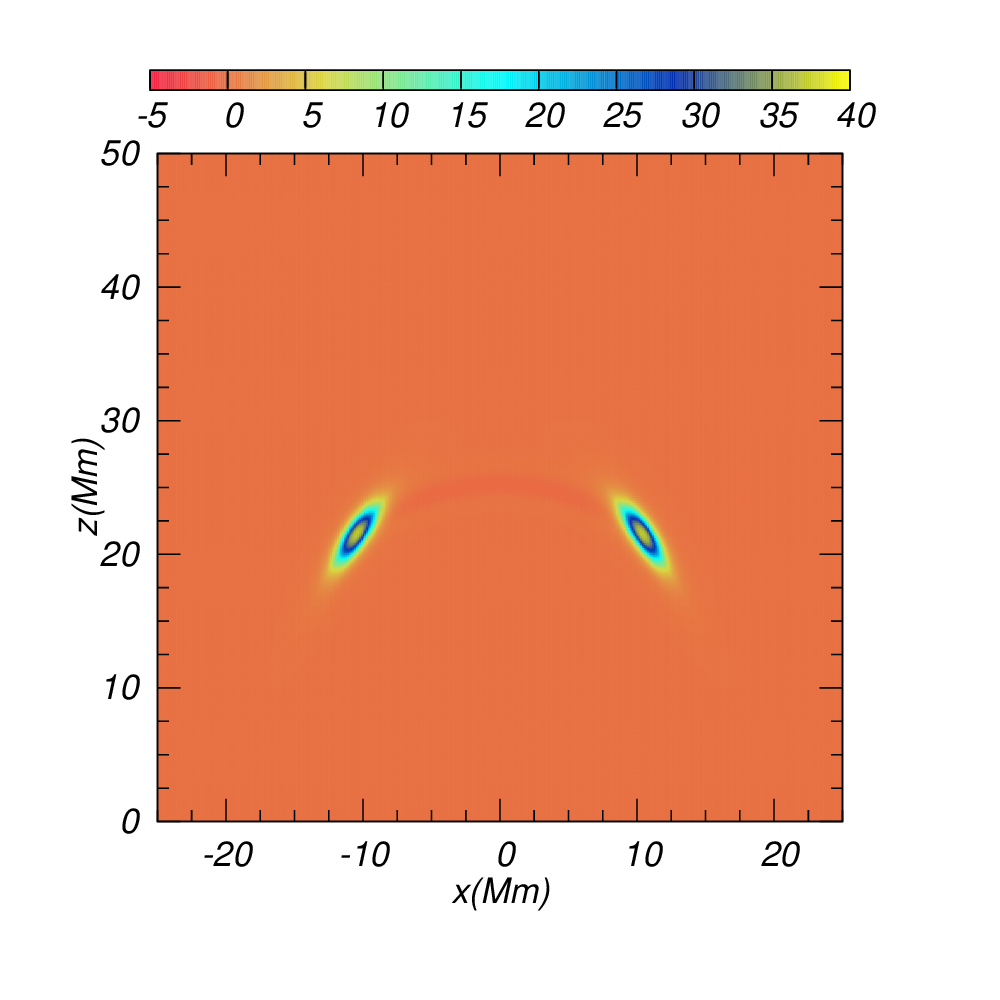}
\includegraphics[width=6.0cm]{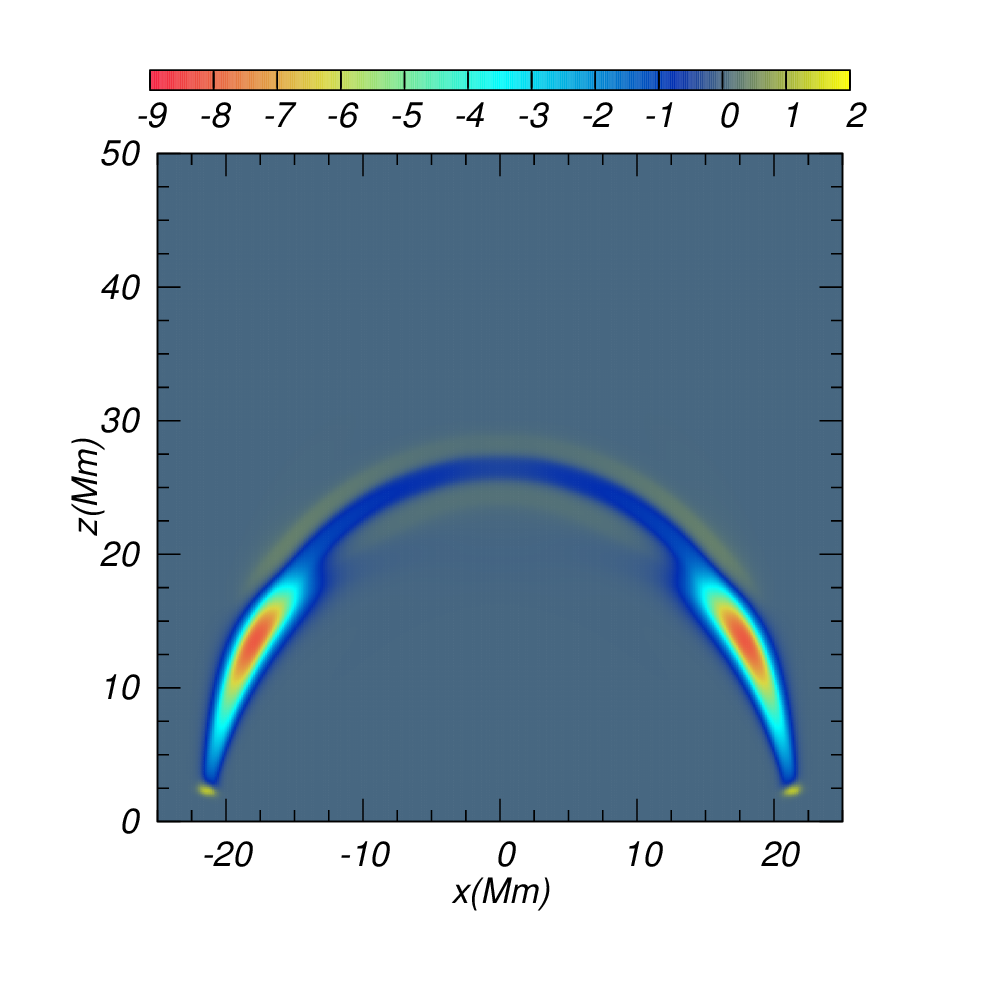}
\includegraphics[width=6.0cm]{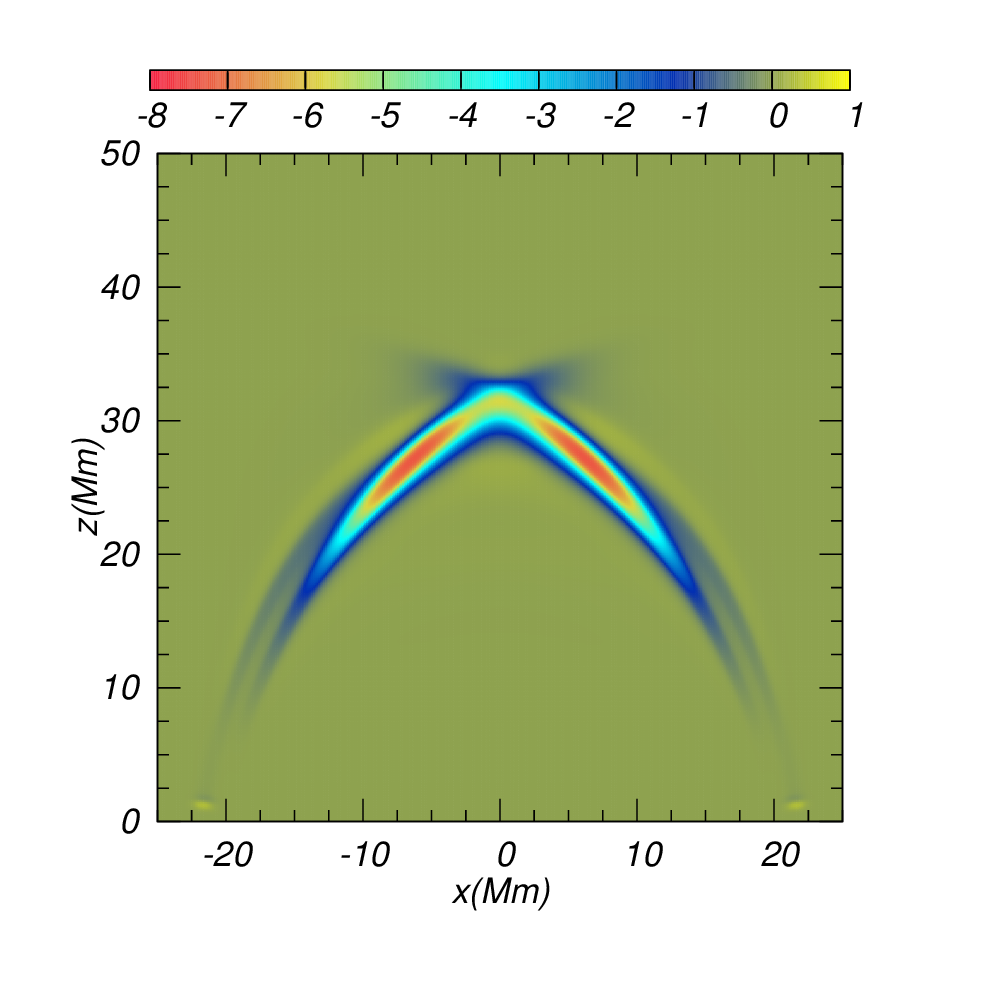}
\includegraphics[width=6.0cm]{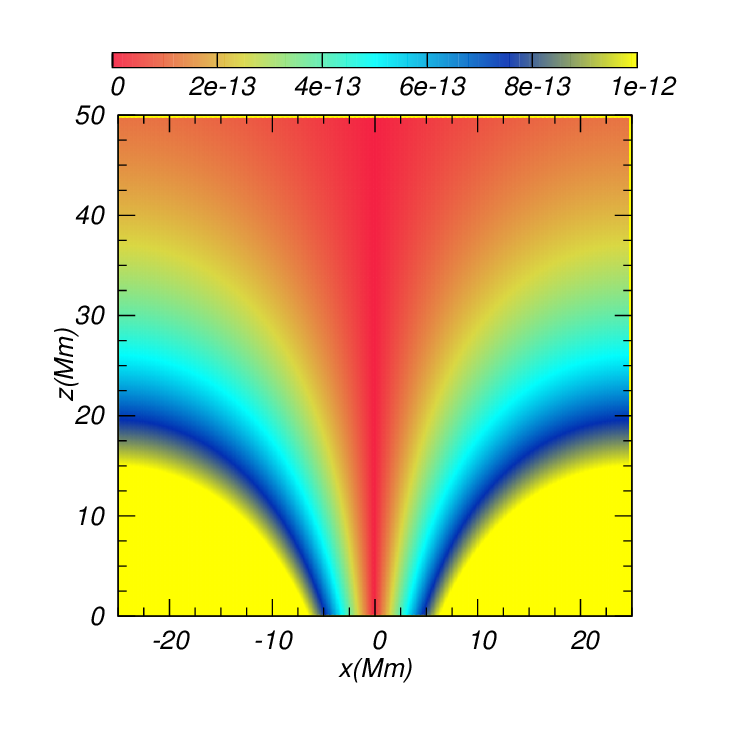}
\caption{\label{fig:tranverse_pulse_symmetric} Transverse velocity $v_y$(km/s) at times $t=10$ s, $t=40$ s, $t=100$ s, $t=200$ s and $\nabla\cdot B$ (Tesla/km) at $t=200$ s.}
\end{center}
\end{figure*}

We consider the case of a symmetric pulse given by the equation (\ref{transverse_pulse}). In Fig. \ref{fig:tranverse_pulse_symmetric} we show snapshots of the transverse velocity $v_y$ at times $t=10$ s, $t=40$ s, $t=100$ s and $t=200$ s. At time $t=10$ s we can see the propagation of two Alfv\'enic pulses with the same amplitude move downward following the fieldlines of coronal loop. At time $t=40$ s the pulses have passed transition region where their amplitude decrease. The pulses move in regions with low density at time $t=100$ s, so the pulses assume a elongated shape. At time $t=200$ s the pulses hit the transition region twice and came back to the center of domain. In this time it is clear the spreading of the pulse, that goes together with a high amplitude decrease produce by reflection at the transition region. Even though the complex dynamics of the Alfv\'enic pulses, the $\nabla\cdot B$ is maintained in the order of $10^{-12}$ Tesla/km as is shown in Fig. \ref{fig:tranverse_pulse_symmetric} (bottom).

%-----> Magnetohydrodynamic-gravity waves ---------------------------------------
\subsubsection{Propagation of magnetohydrodynamic-gravity waves in the 3D solar atmosphere}

Following \cite{Murawski_et_al_2013}, we simulate the evolution magnetohydrodynamic-gravity waves and the vortex formation in the magnetized three-dimensional (3D) solar atmosphere. Initially we assume the solar atmosphere to be in static equilibrium \cite{Murawski&Musielak_2010,Murawski&Zaqarashvili_2010,Murawski_et_al_2011,Woloszkiewicz_et_al_2014,Chmielewski_et_al_2014a,Chmielewski_et_al_2014b,Gonzalez_Aviles_Guzman_2015}, and the initial equilibrium magnetic field ${\bf B_e}$ constant along the vertical direction

\begin{equation}
{\bf B_e}= B_{0}{\bf\hat{z}},
\end{equation}

\noindent where $B_{0}=23$ G, which is a value close to a typical magnetic field magnitude in the quiet Sun \cite{Vogler&Schussler_2007}. 

As a result of applying the static equilibrium and the magnetic current-free conditions to equation (\ref{momentum}),  the pressure gradient is balanced by gravity through equation (\ref{equilibrium_pressure}). In addition, considering the fact that hydrostatic equilibrium will only hold along the $z$ direction, the equilibrium gas pressure takes the form:

\begin{equation}
p_{e}(z) = p_{ref}\exp\left(-\int_{z_{ref}}^{z}\frac{dz^{\prime}}{\Lambda(z^{\prime})}\right), \label{pe_profile}
\end{equation}

\noindent and the mass density

\begin{equation}
\rho_{e}(z) = \frac{p_{e}(z)}{g\Lambda(z)}, \label{rhoe_profile} 
\end{equation}

\noindent where

\begin{equation}
\Lambda(z) = \frac{k_{B}T_{e}(z)}{mg} \label{scale_pressure}, 
\end{equation}

\noindent is the pressure scale-height, and $p_{ref}$ represents the gas pressure at a given reference level, that we choose to be located at $z_{ref}=10$ Mm, because in the temperature model above, this is the location at which the corona region starts. In this case we use $\mu=$1.24 in (\ref{eos_temp}) only to reproduce the results in \cite{Murawski_et_al_2013}.

\begin{figure}
\begin{center}
\includegraphics[width=5.0cm]{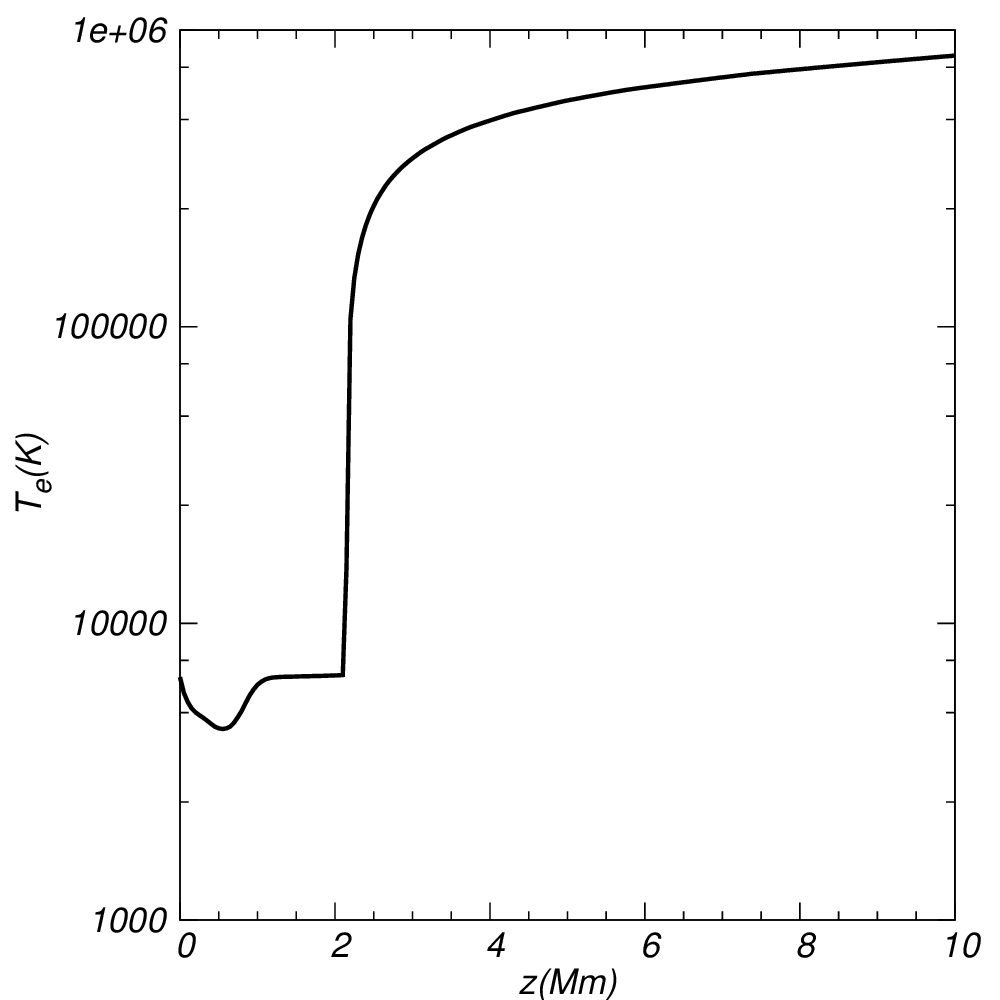}
\includegraphics[width=5.0cm]{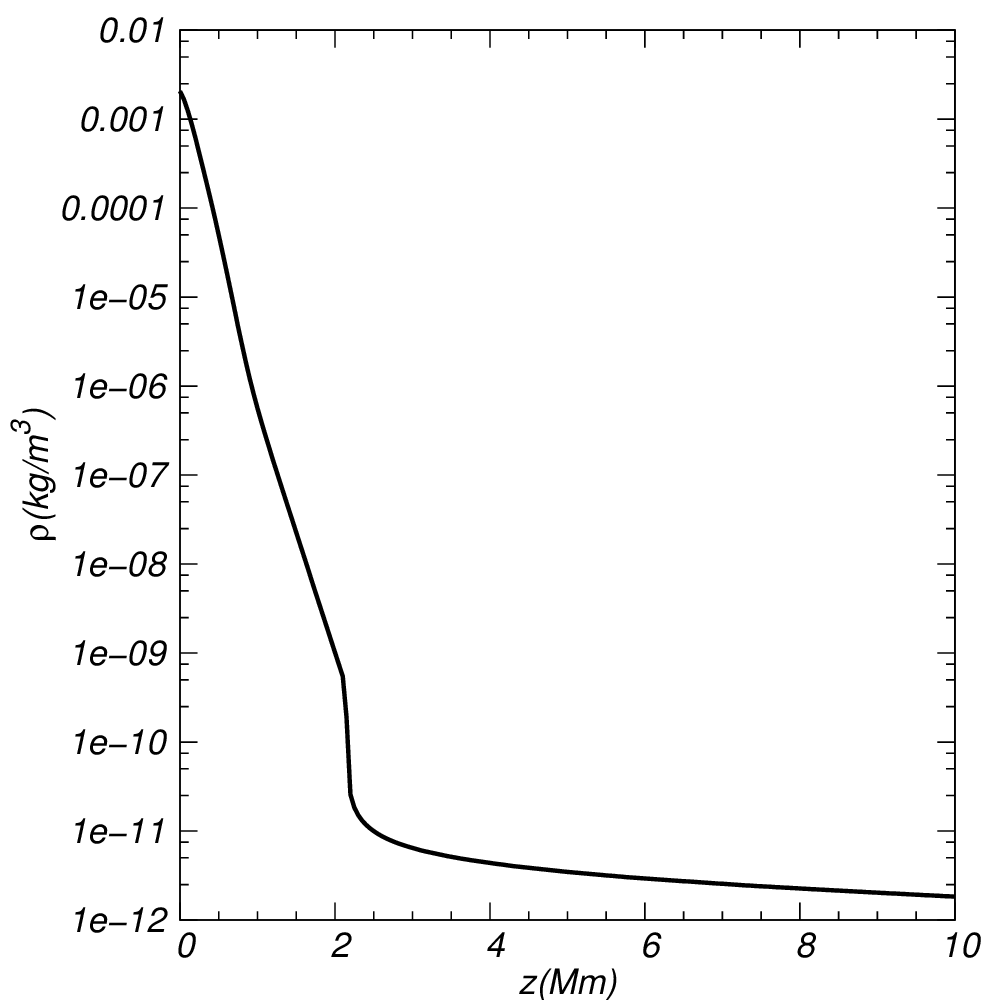}
\includegraphics[width=5.0cm]{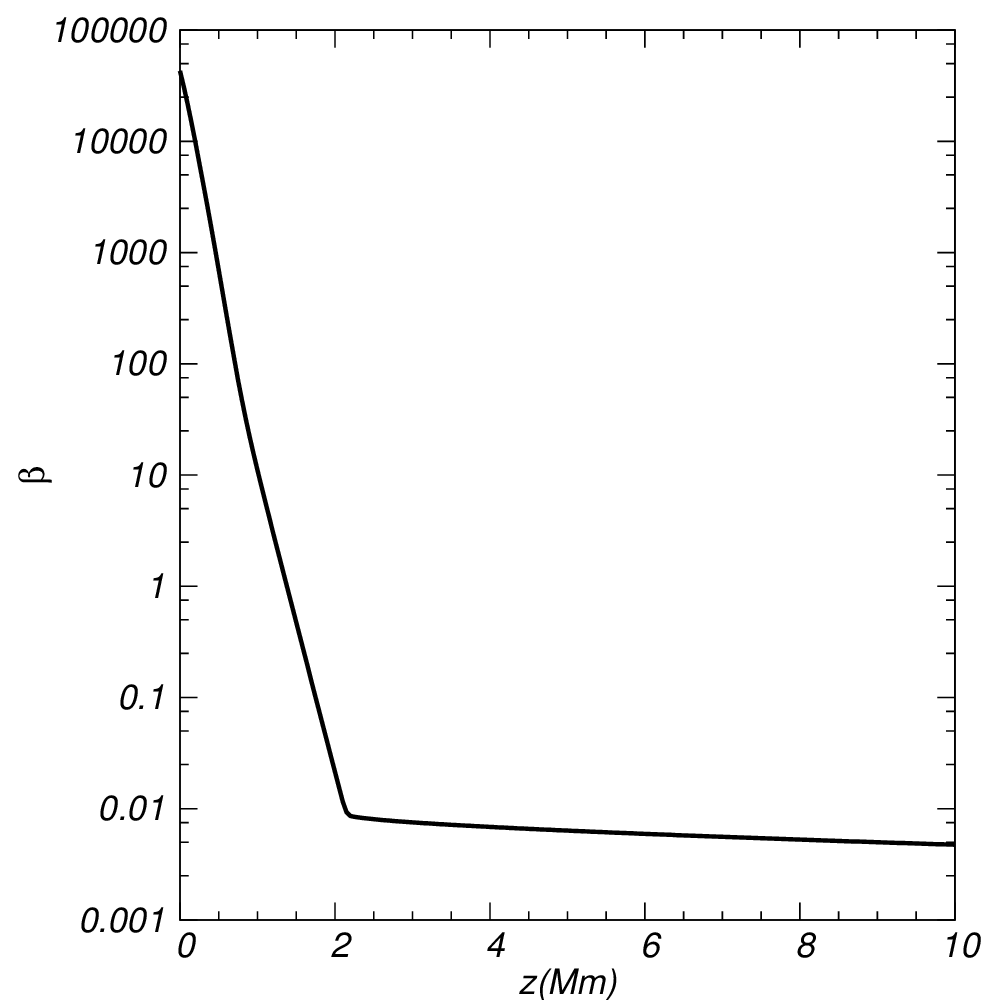}
\caption{\label{fig:equilibrium_profiles} Temperature, density and $\beta$ profile for the C7 model.}
\end{center}
\end{figure}

The temperature field $T_{e}(z)$ is assumed to obey the C7 model \cite{Avrett&Loeser2008}, which is a semiempirical model of the chromosphere. This C7 temperature profile was interpolated into the 3D numerical grid, then we numerically integrated equation (\ref{pe_profile}) in order to obtain $p_e(z)$ and finally with this profile we use equation (\ref{rhoe_profile}) to obtain $\rho_e(z)$. We show the temperature and density profile in Fig. \ref{fig:equilibrium_profiles} where the important density gradients can be observed. In addition we show the plasma $\beta$ to see how the magnetic pressure dominates over gas pressure in the solar atmosphere.

In order to simulate the propagation of magnetohydrodynamic-gravity waves, we perturb the static solar atmosphere with a Gaussian pulse, either in the horizontal component of velocity, $v_{x}$, or in the vertical component $v_z$. Following \cite{Murawski_et_al_2013}, the initial conditions are given by

\begin{eqnarray}
v_{x}(x,y,z,t=0) &=& A_{v}e^{-\frac{x^{2}+y^{2}+(z-z_0)^{2}}{w^{2}}}, \label{vx_perturbation} \\
v_{z}(x,y,z,t=0) &=& A_{v}e^{-\frac{x^{2}+y^{2}+(z-z_0)^{2}}{w^{2}}}, \label{vz_perturbation}
\end{eqnarray}

\noindent where $A_v$ is the amplitude of the pulse, $z_0$ is the inital position over the vertical direction and $w$ denotes the width of the pulse. In this case we use the values $A_v=3$ kms$^{-1}$, $w=100$ km and $z_0=500$ km. Therefore the initial conditions for this problem are given by: $p=p_e(z)$, $\rho=\rho_e(z)$, the velocity field ${\bf v}=(v_x,0,v_z)$ depends on the direction of perturbation, and ${\bf B}=(0,0,B_0)$.  

We solve the equations (\ref{density})-(\ref{divergenceB}) using the HLLE Riemann solver, the Minmod slope limiter, a Courant factor $CFL$=0.2 and $\gamma=1.4$. We set the simulation domain to [-0.75,0.75]$\times$[0,3]$\times$[-0.75,0.75] Mm$^{3}$, covered with 64$\times$128$\times$64 cells for the case of horizontal perturbation and 100$\times$200$\times$100 cells for vertical perturbation. At the top and bottom faces of the numerical domain we impose fixed in time boundary conditions, whereas at the remaining four sides we apply open boundary conditions. In our simulations we choose the scale factors in cgs units as specified in Table \ref{table:1}. These units are obtained choosing a length-scale $l_0$, plasma density scale $\rho_0$ and magnetic field scale $B_0$, this fixes the unit of time in terms of Alfv\'en speed as follows:

\begin{equation}
B_0 = v_0\sqrt{\mu_0\rho_0},
\end{equation}

\noindent where $\mu_0=4\pi\times10^{-7}$ $N/A^{2}$ is the magnetic permeability and $\rho_0=10^{-12}$ $kg/m^{3}$ is the plasma scale at the level of the solar corona.

\begin{table}
\caption{Units used in this paper}
% title of Table
\centering
% used for centering table
\begin{tabular}{c c c c}
% centered columns (4 columns)
\hline\hline
%inserts double horizontal lines
Quantity & cgs units \\ [0.5ex]
% inserts table
%heading
\hline
% inserts single horizontal line
$l_{0}$ & $10^{8}$ cm \\
$v_{0}$ & $10^{8}$ cm$\cdot s^{-1}$ \\
$t_{0}$ & 1 s \\
$\rho_{0}$ & $10^{-15}$ gr$\cdot cm^{-3}$ \\
$B_{0}$ & 11.21 G \\ [1ex]
% [1ex] adds vertical space
\hline
%inserts single line
\end{tabular}
\label{table:1}
% is used to refer this table in the text
\end{table}

%----------Vertical perturbation --------------------------
\subsubsection{Vertical perturbation}

\begin{figure*}
 \begin{tabular}{ccc} \hline
  $t=37.5$ s & $t=100$ s & $t=150$ s \\ \hline
  && \\ 
 \multicolumn{3}{c}{$v_z(km/s)$ in $yz$ plane}\\ 
     \includegraphics[width=5.3cm]{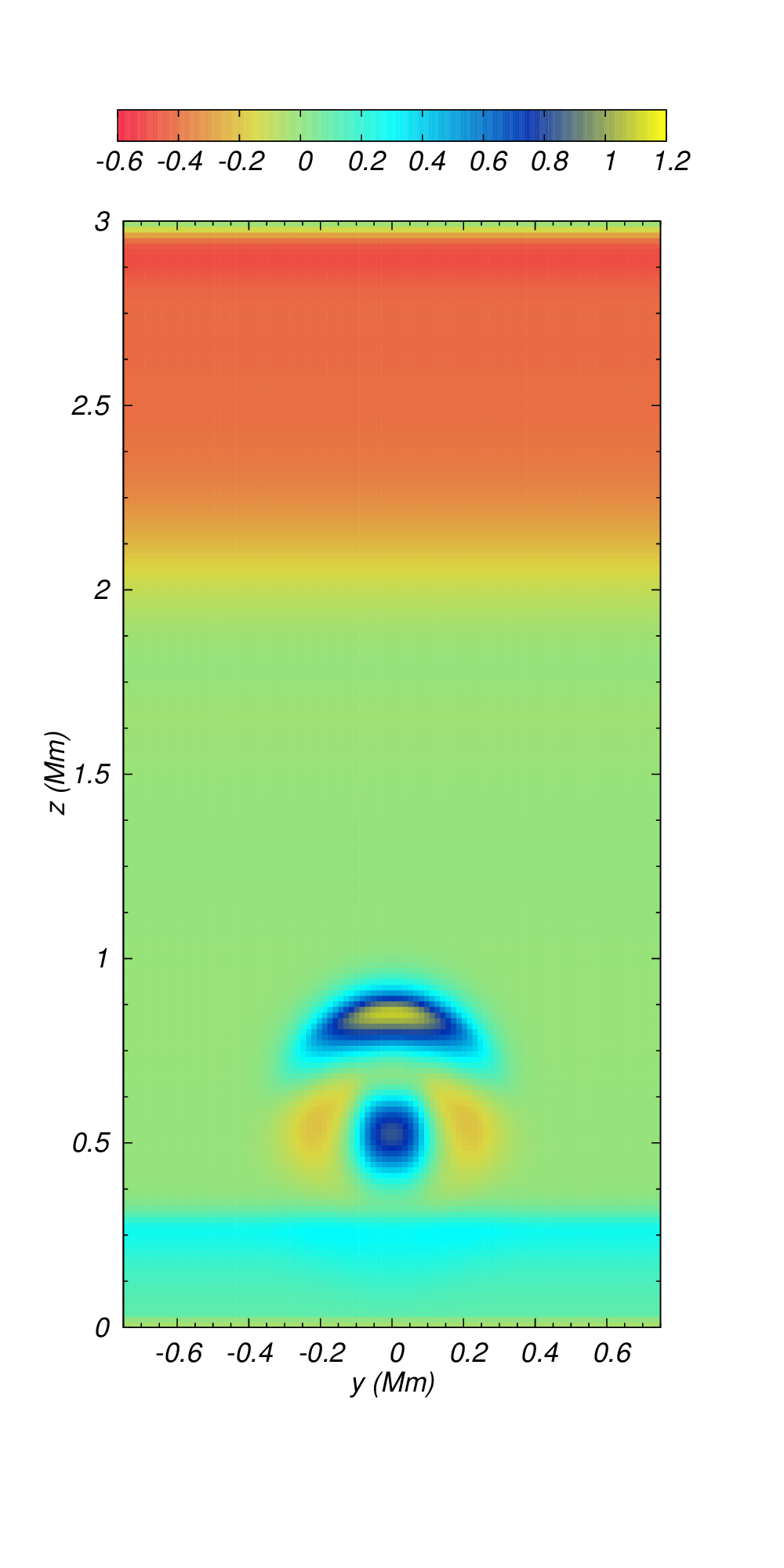}&
     \includegraphics[width=5.3cm]{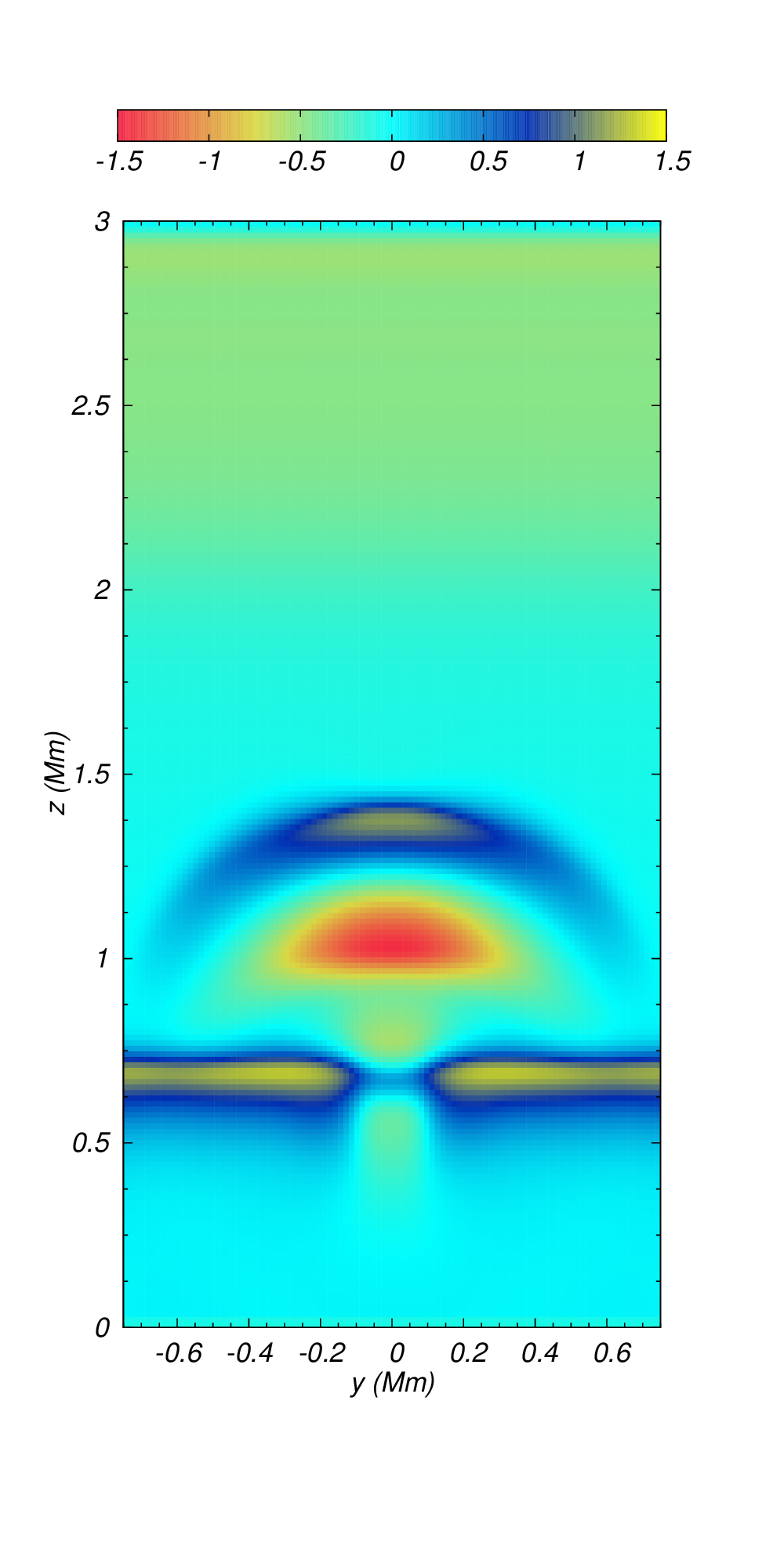}&
     \includegraphics[width=5.3cm]{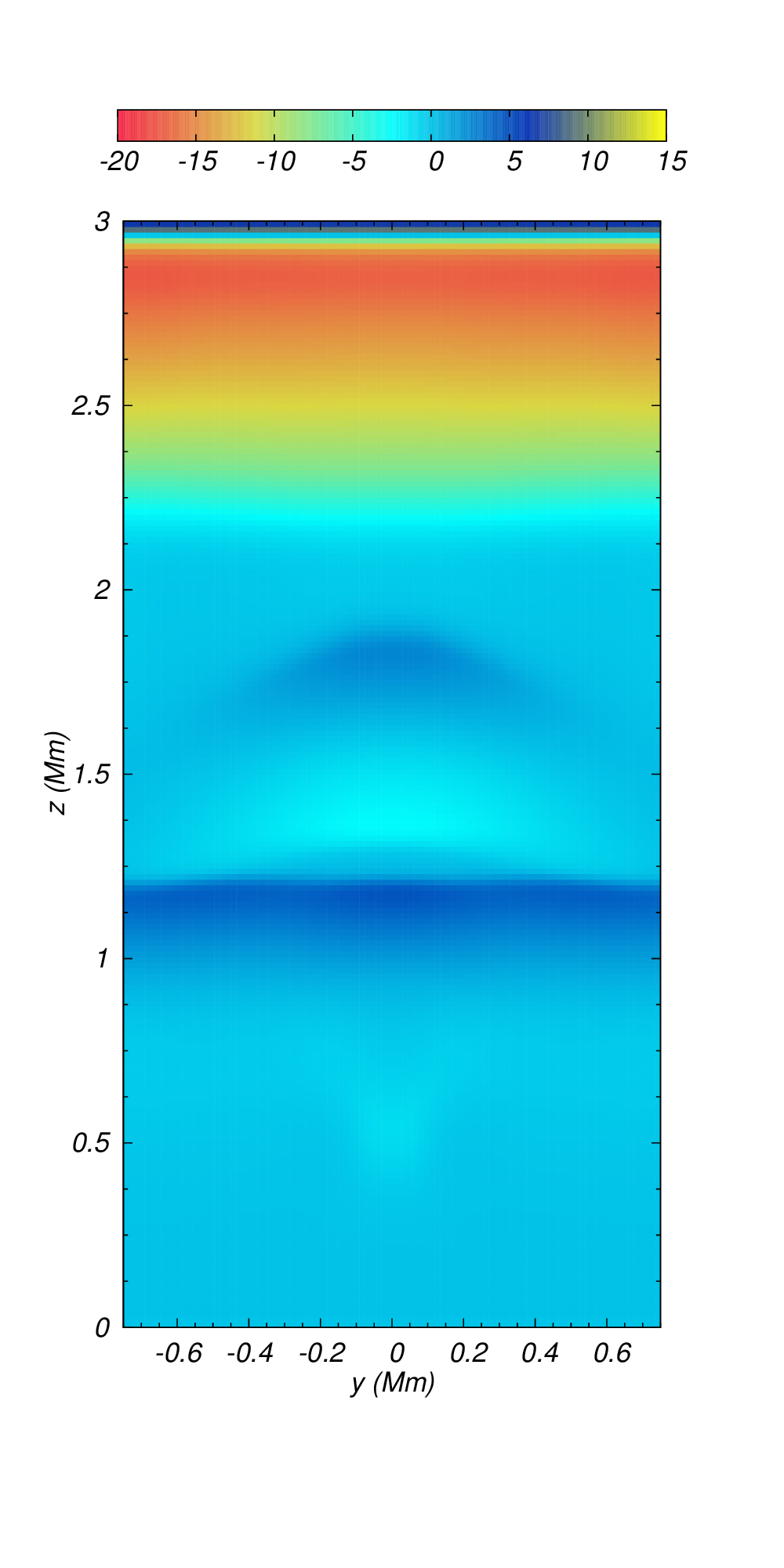}\\
 \multicolumn{3}{c}{$v_z(km/s)$ in $xz$ plane}\\ 
     \includegraphics[width=5.3cm]{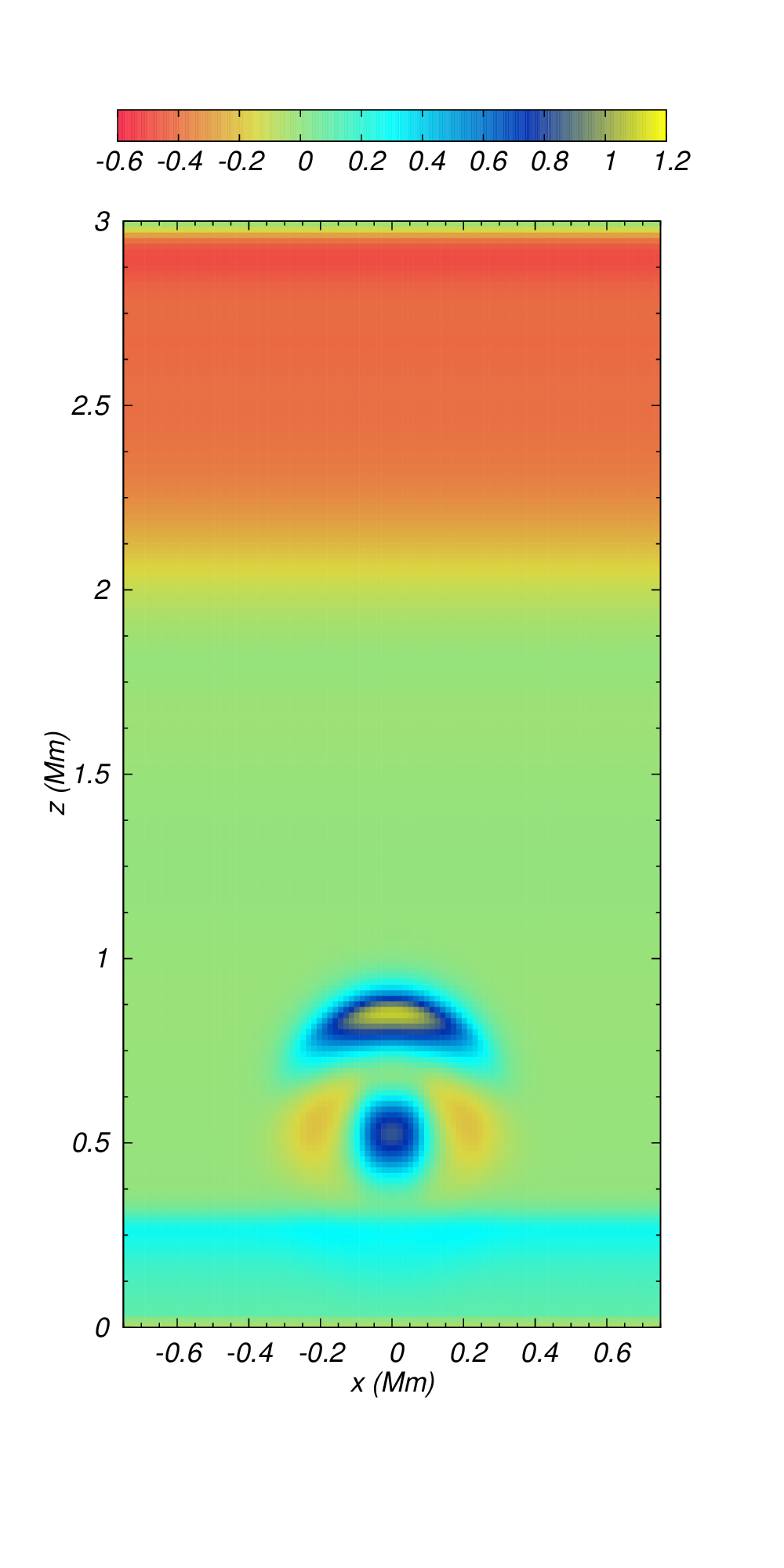}&
     \includegraphics[width=5.3cm]{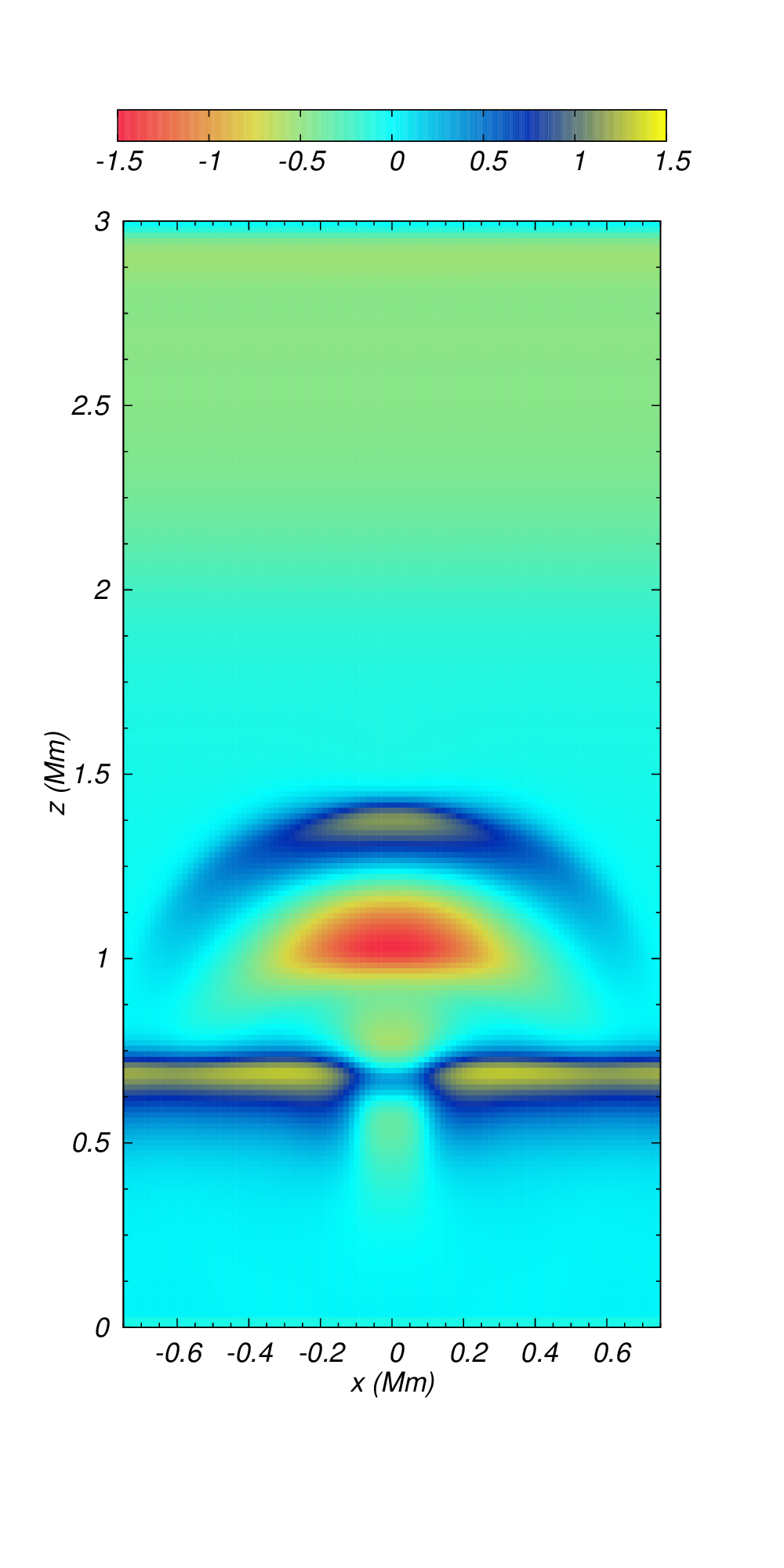}&
     \includegraphics[width=5.3cm]{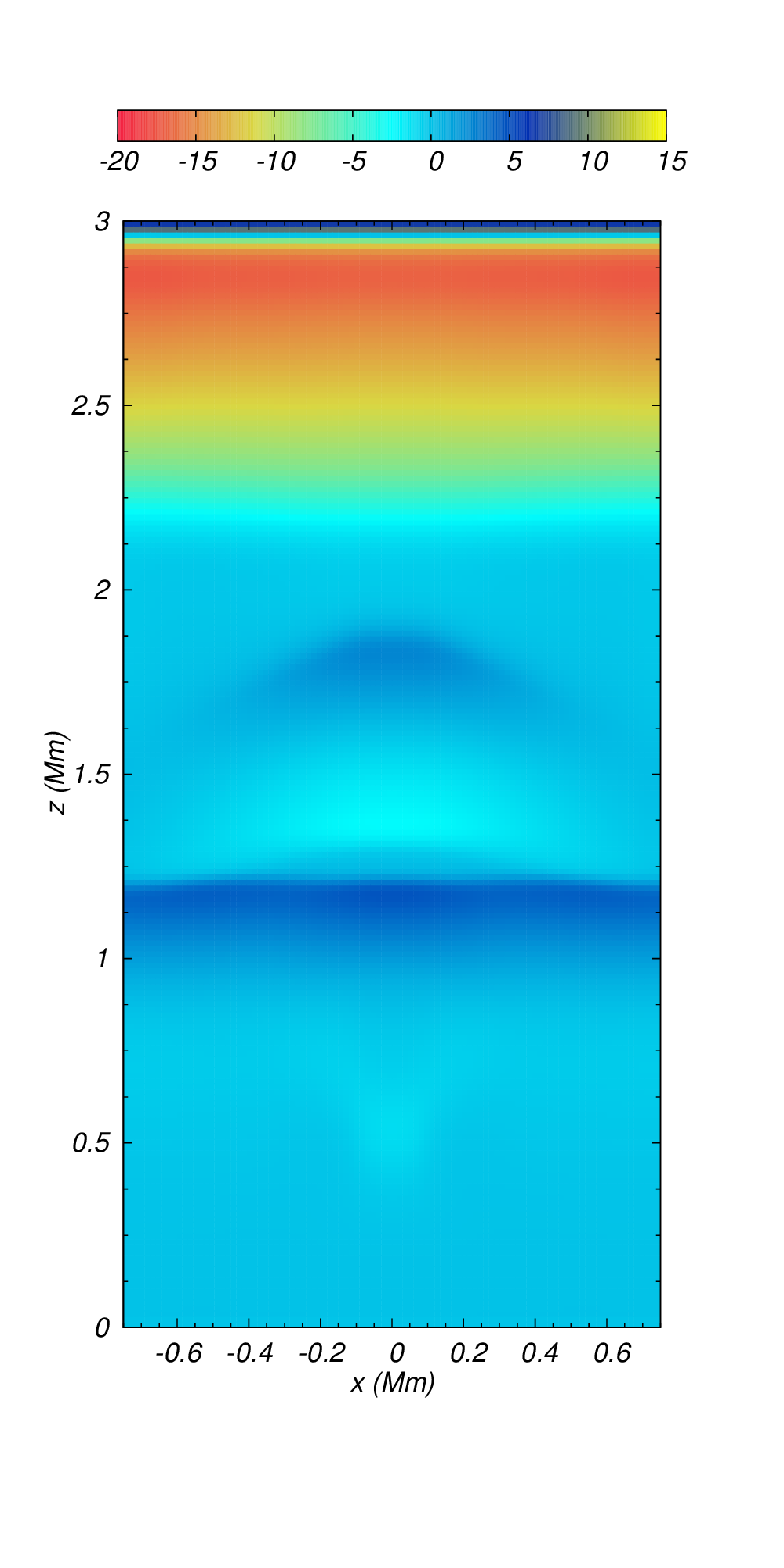}\\
 \hline
   \end{tabular}
   \caption{\label{fig:vertical_perturbation_yz_xz} Snapshots of the vertical component of velocity $v_z(km/s)$ at times $t=37.5,100,150$ s in the $yz$ and $xz$ planes.}
   \end{figure*}

\begin{figure*}
  \begin{tabular}{ccc} \hline
  $t=37.5$ s & $t=100$ s & $t=150$ s \\ \hline
  && \\ 
 \multicolumn{3}{c}{$v_z(km/s)$ in $xy$ plane}\\ 
     \includegraphics[width=6.0cm]{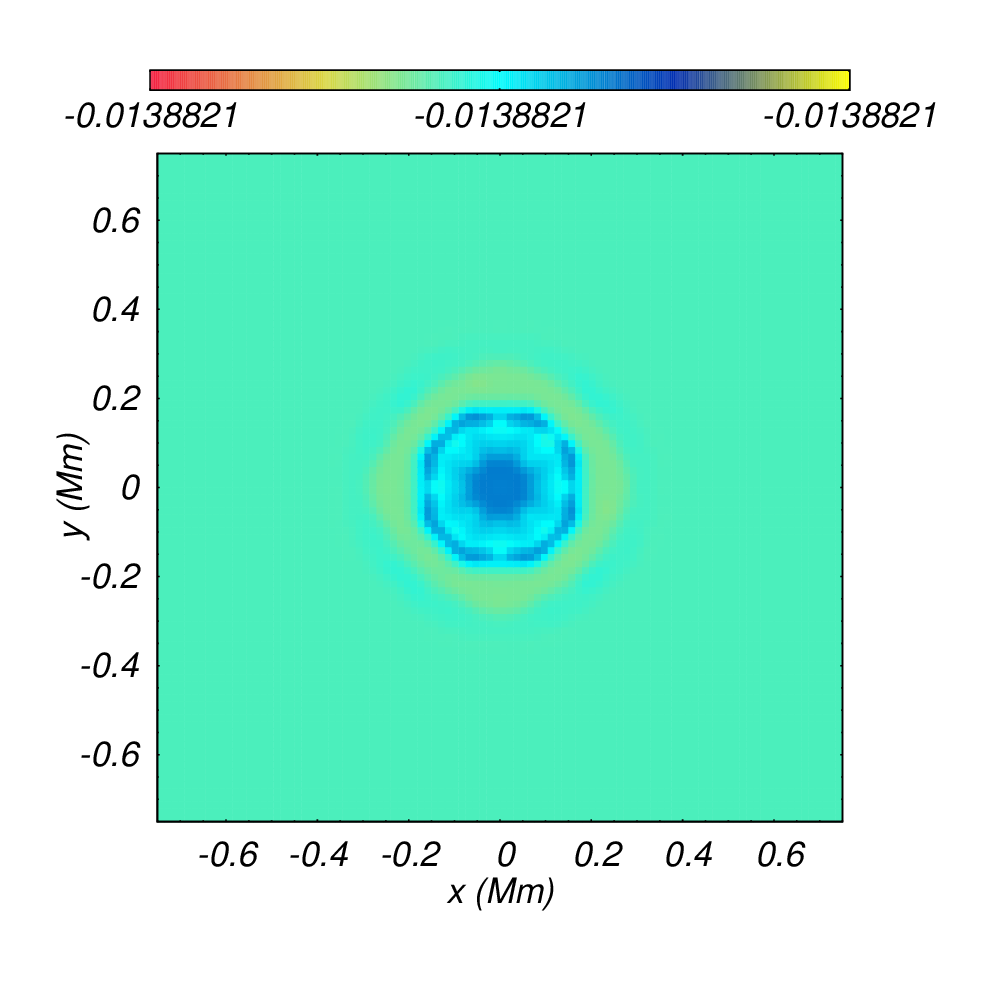}&
     \includegraphics[width=6.0cm]{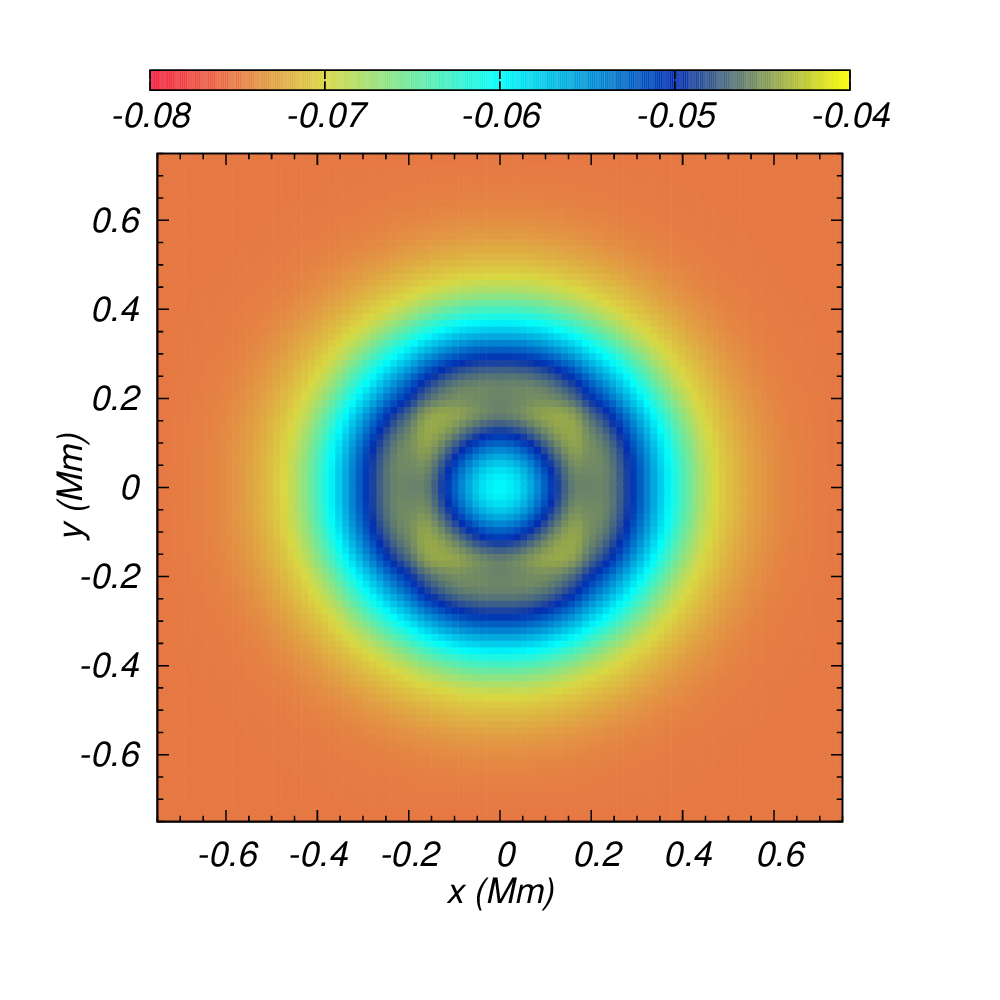}&
     \includegraphics[width=6.0cm]{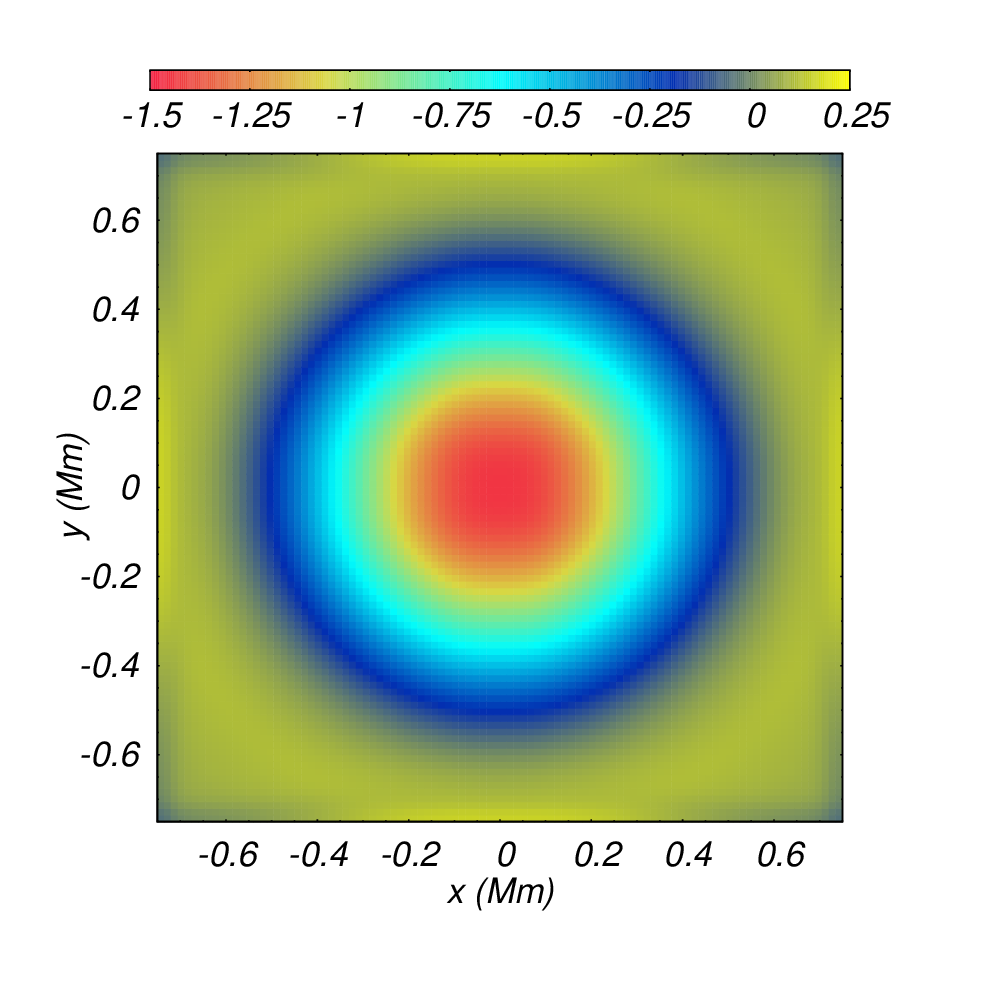}\\
   \multicolumn{3}{c}{$|\nabla\cdot B|(Tesla/km)$ in $xy$ plane}\\ 
     &
     \includegraphics[width=6.0cm]{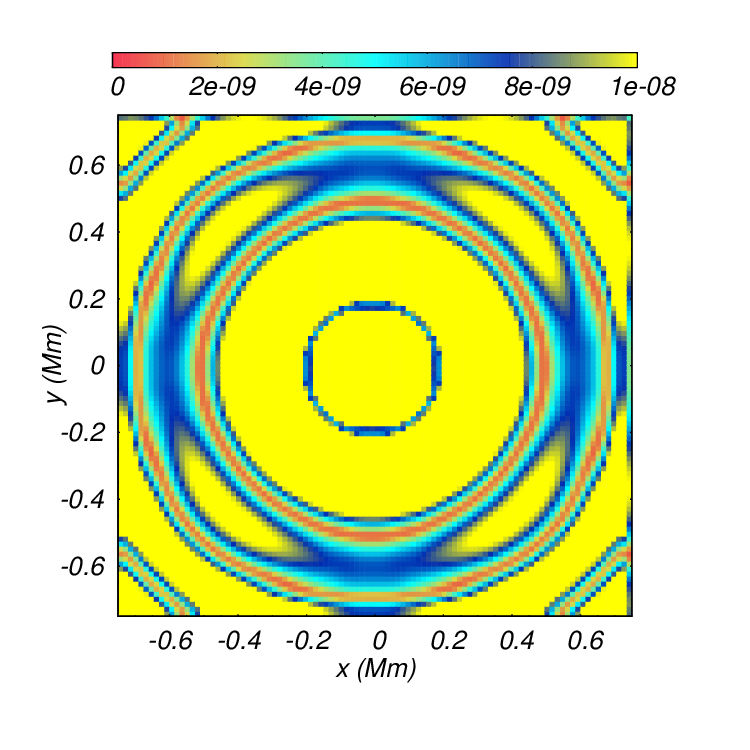}\\     
 \hline
   \end{tabular}
   \caption{\label{fig:vertical_perturbation_xy} Snapshots of the vertical component of velocity $v_z(km/s)$ at times $t=37.5,100,150$ s and 
$|\nabla\cdot B|$ at $t=150$ s in the $xy$ plane.}
  \end{figure*}

Vertical perturbations in the sun could be generated by $p$-modes \cite{Jain_et_al_2011}, which are pressure or acoustic modes with frequencies greater than 1 mHz. This kind of perturbation is given by the component $v_z$ in equation (\ref{vz_perturbation}). The initial pulse triggers longitudinal magnetoacoustic-gravity waves. At the initial point where the wave is launched, the gas pressure is greater than the magnetic pressure, which implies that  $\beta>1$. This excites fast and slow magnetoacoustic-gravity waves which are coupled. The initial pulse splits into two propagating waves at time $t=37.5$ s shown in Fig. \ref{fig:vertical_perturbation_yz_xz} (top). At time $t=100$ s, the two magnetocoustic waves move upwards until $z\approx1.4$ Mm, in the same way as in Fig. 3 of \cite{Murawski_et_al_2013}. The waves continue moving until they reach the transition region. This vertical perturbation produces the propagation of magnetoacoustic-gravity waves in a symmetric way, as shown in Fig. \ref{fig:vertical_perturbation_yz_xz} at the $xz$ and $yz$ planes. In these two planes the propagation is seen in the same way. We also show the propagation as seen from above, on the $xy$  plane in Fig. \ref{fig:vertical_perturbation_xy}. Even though the structure of the magnetoacoustic-gravity waves in the solar atmosphere looks complex, the value of $|\nabla\cdot B|$ at time $t=150$ s in the $xy$ plane remains low as is shown in Fig. \ref{fig:vertical_perturbation_xy}.

\begin{figure}
\begin{center}
\includegraphics[width=6.0cm]{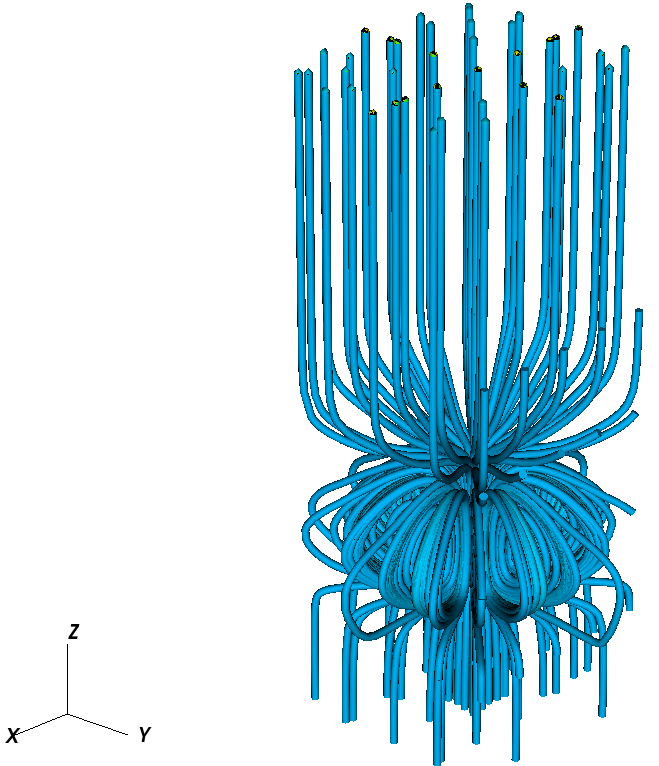}
\includegraphics[width=6.0cm]{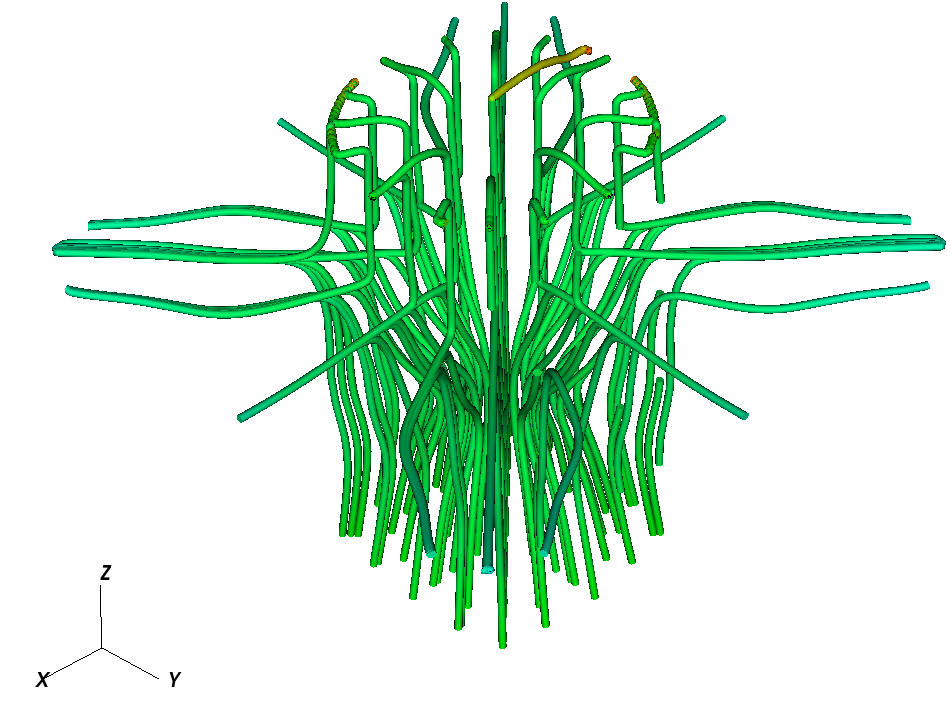}
\caption{\label{fig:streamlines_vertical_pert} Temporal snapshots of streamlines at $t=37.5$ s (top) and $t=150$ s (bottom) for the case of vertical perturbation. Color represents the vorticity magnitude.}
\end{center}
\end{figure}

We show snapshots of the fluid streamlines at times $t=37.5$ s and $t=150$ in Fig. \ref{fig:streamlines_vertical_pert}. These streamlines reveal vortices specially at time $t=37.5$ s, which  are produced by the vertical perturbation $v_z$ and are located at 1 Mm over the photosphere.

%----------Horizontal perturbation --------------------------
\subsubsection{Horizontal perturbation}

\begin{figure*}
 \begin{tabular}{ccc} \hline
  $t=50$ s & $t=150$ s & $t=250$ s \\ \hline
  && \\ 
 \multicolumn{3}{c}{$v_x(km/s)$ in $xz$ plane}\\ 
     \includegraphics[width=5.3cm]{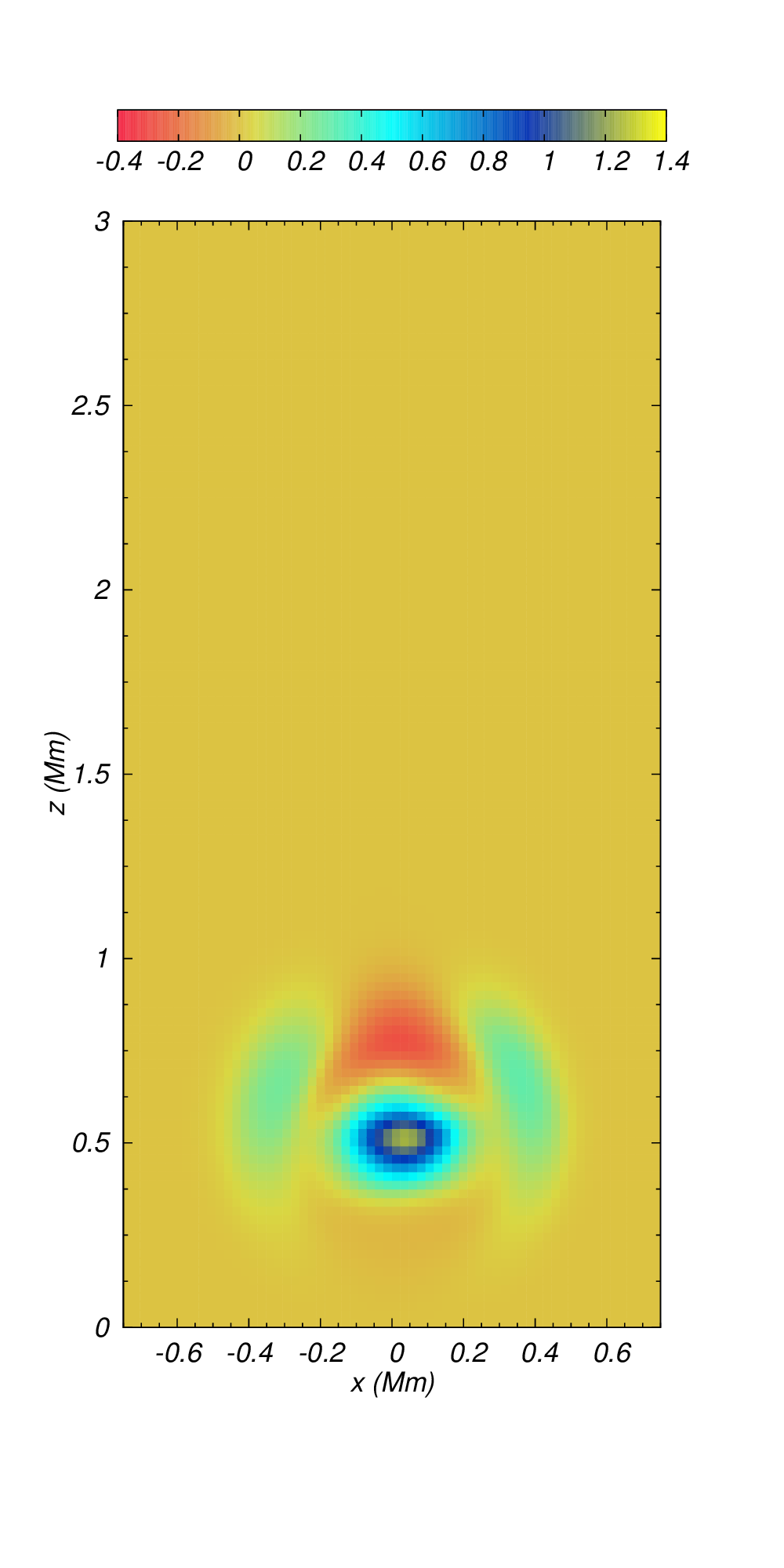}&
     \includegraphics[width=5.3cm]{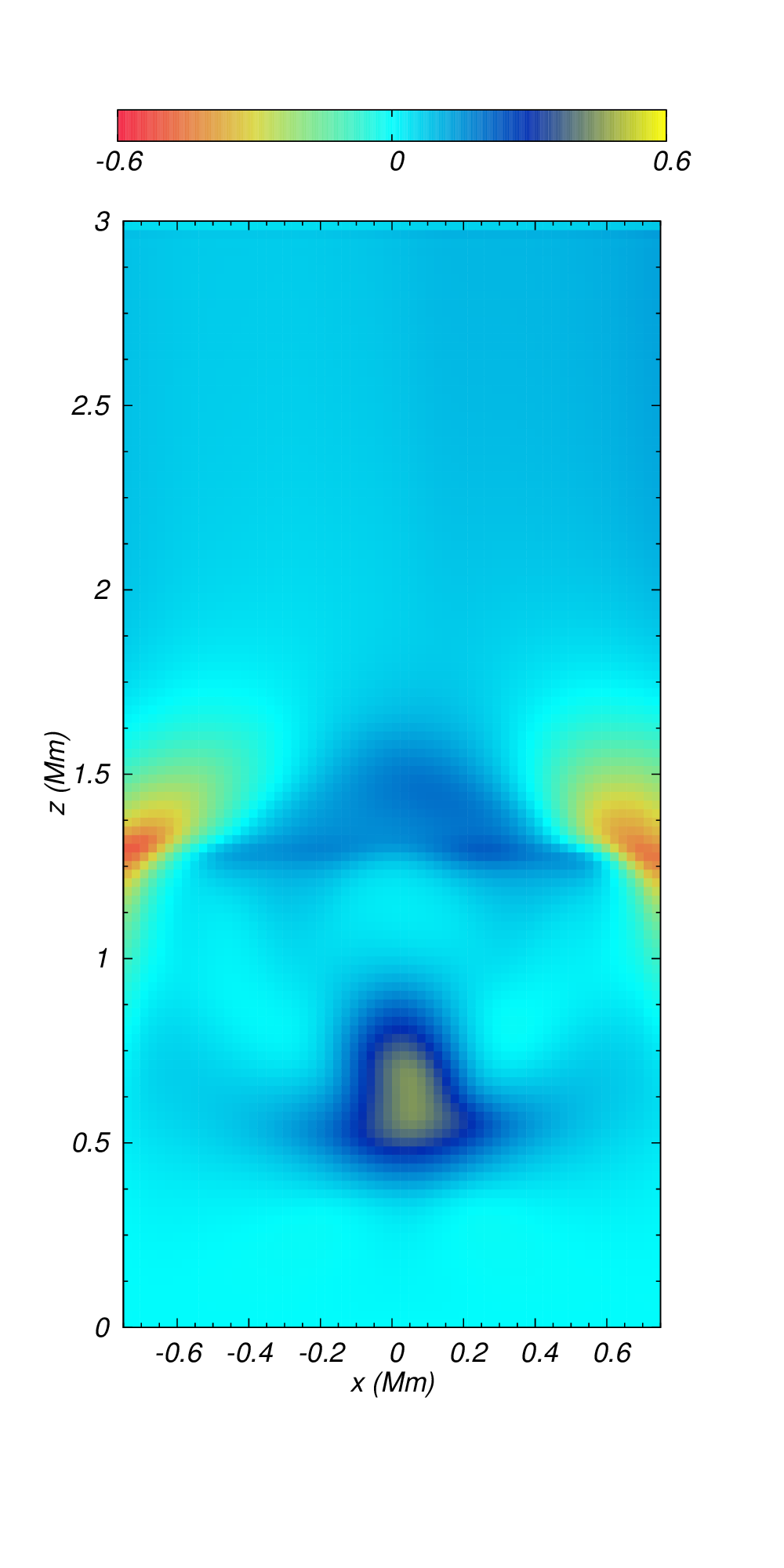}&
     \includegraphics[width=5.3cm]{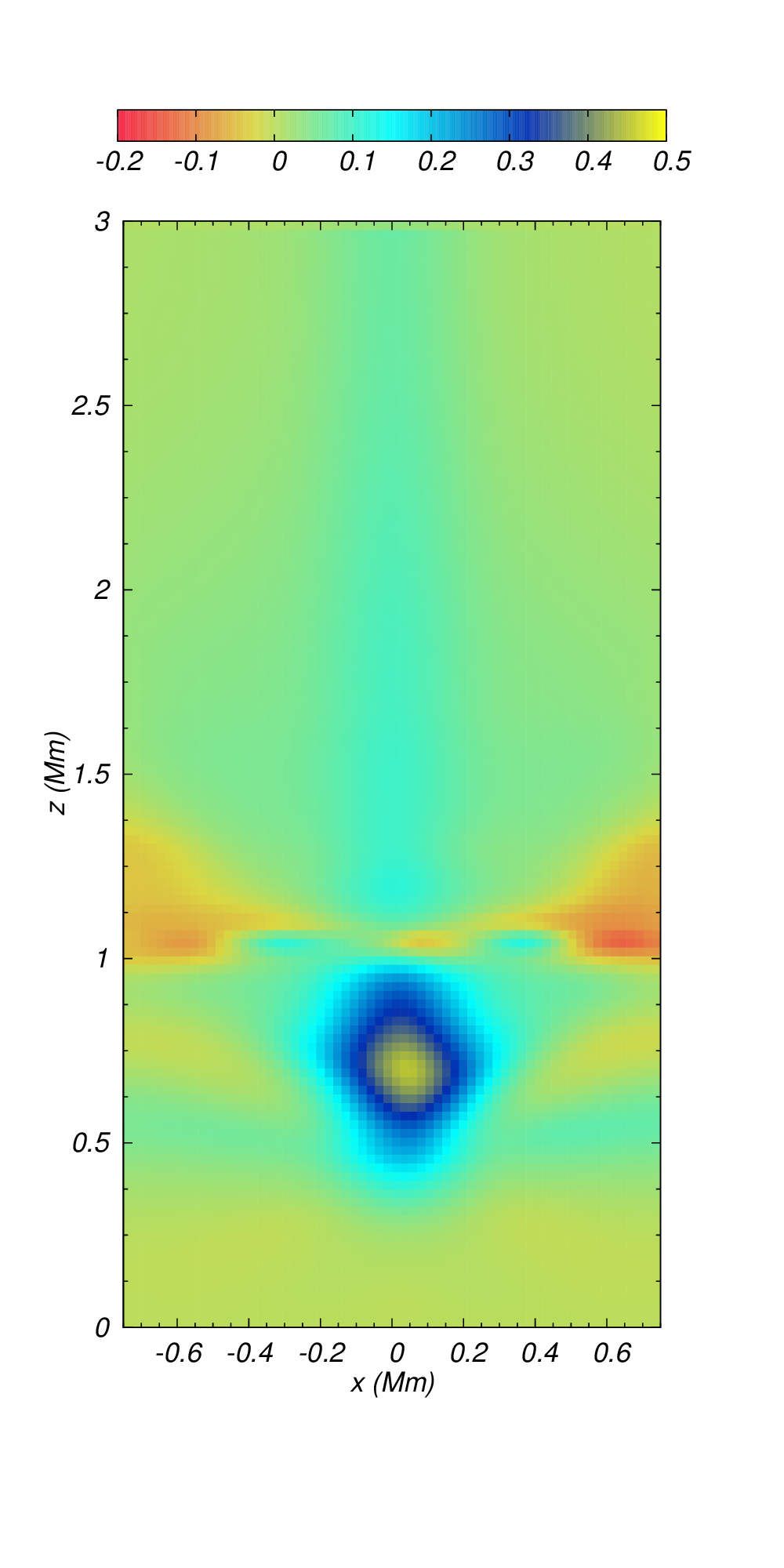}\\
 \multicolumn{3}{c}{$v_x(km/s)$ in $yz$ plane}\\ 
     \includegraphics[width=5.3cm]{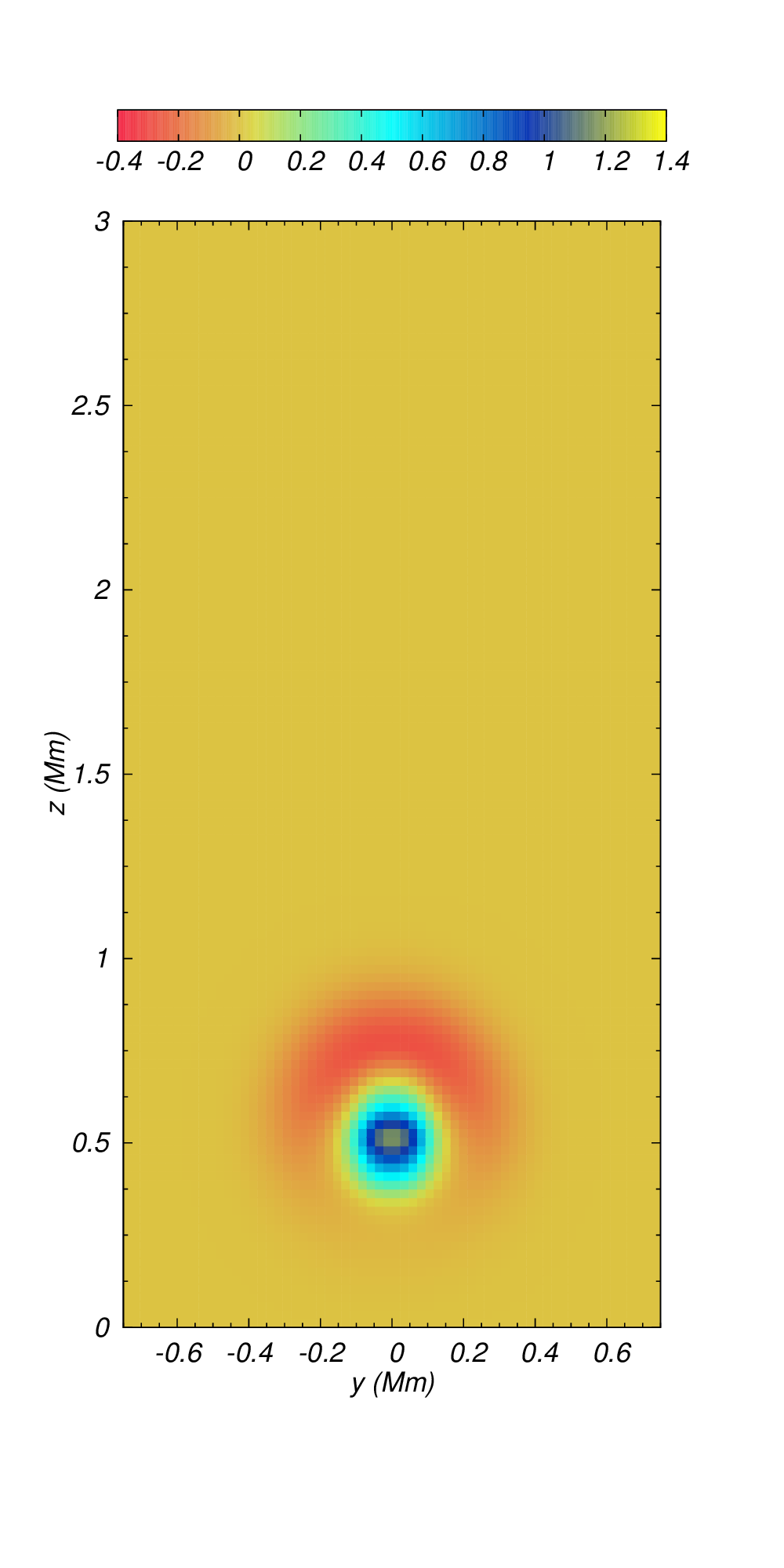}&
     \includegraphics[width=5.3cm]{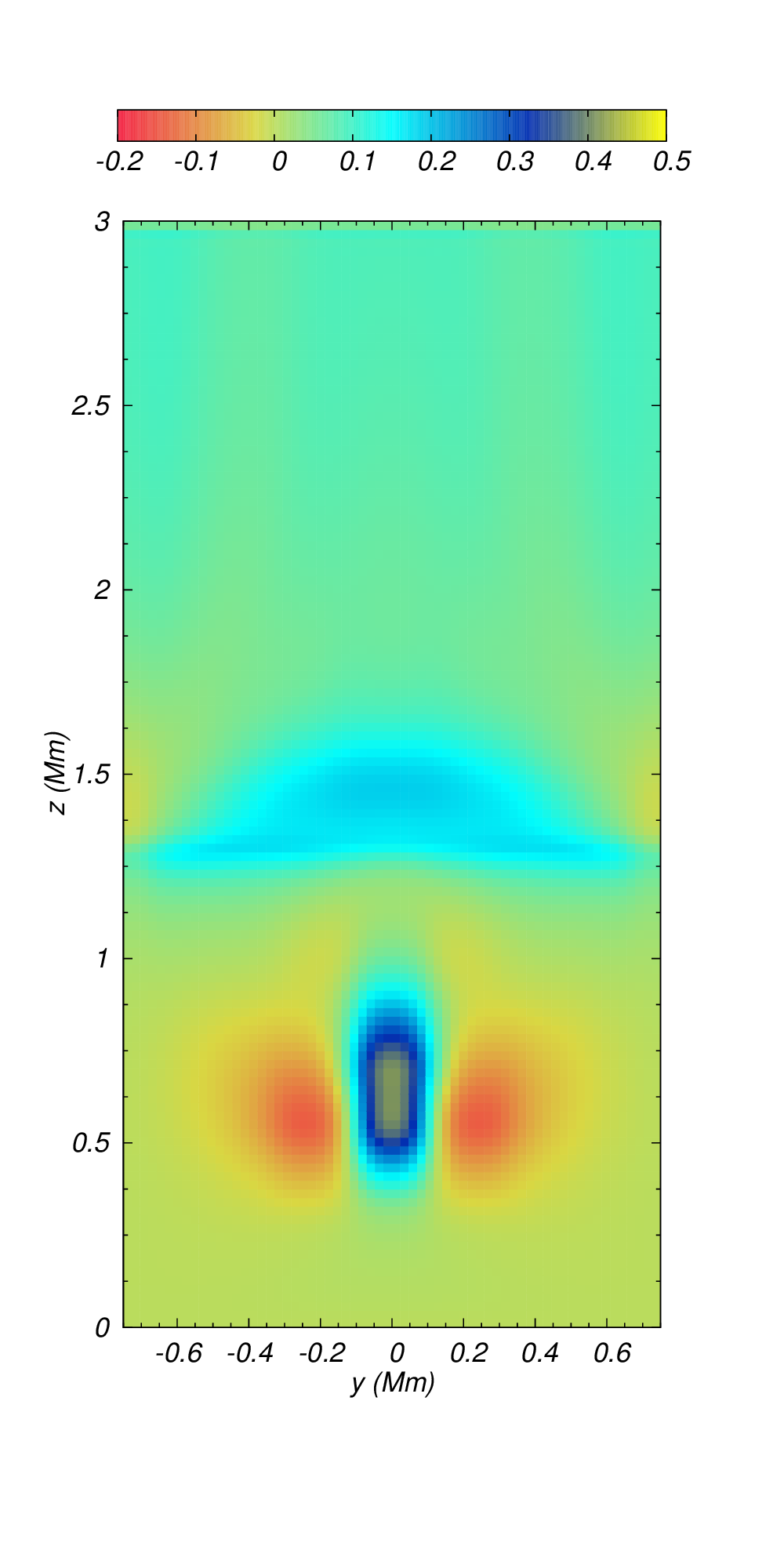}&
     \includegraphics[width=5.3cm]{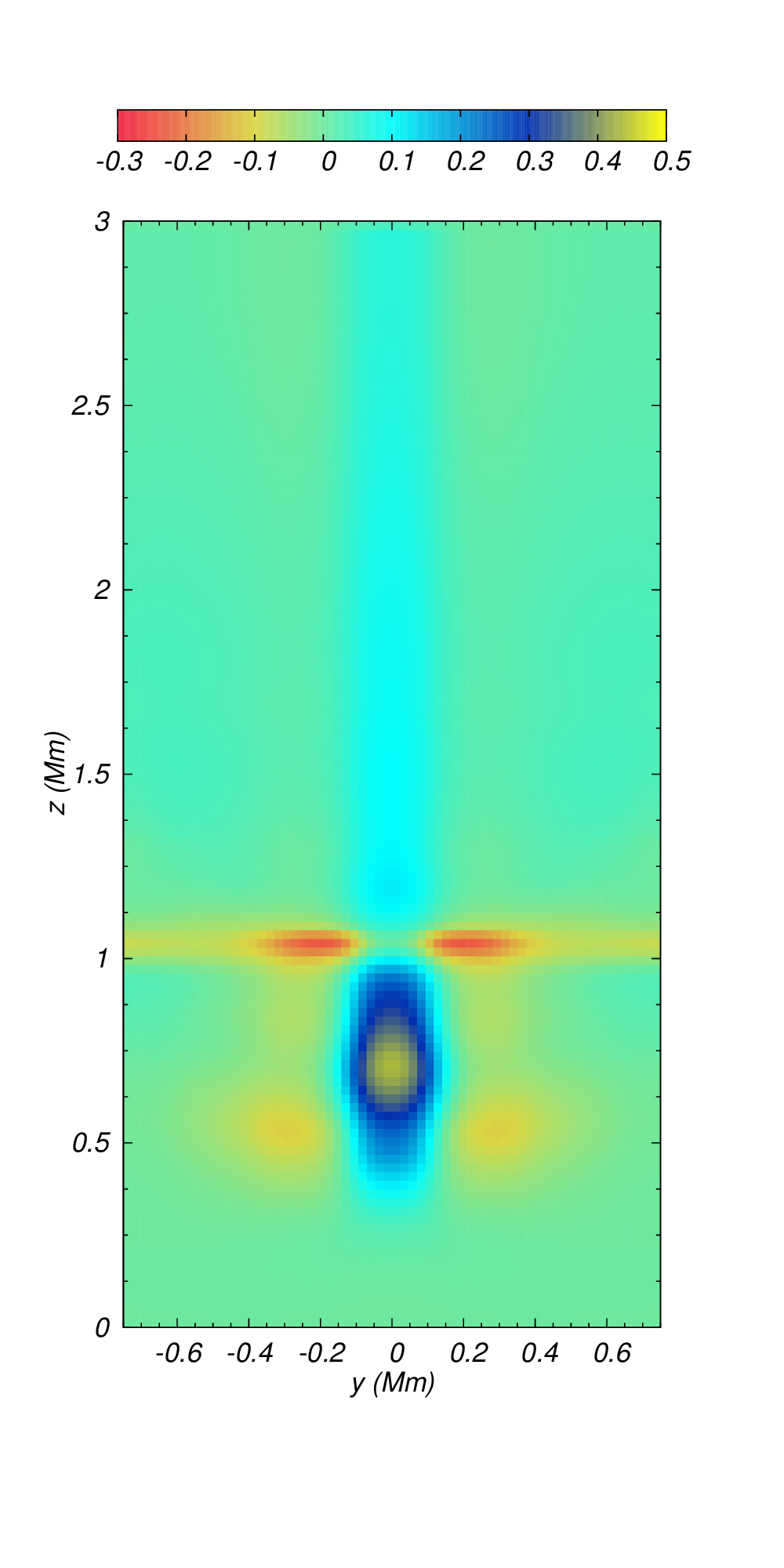}\\
 \hline
   \end{tabular}
   \caption{\label{fig:horizontal_perturbation_yz_xz} Snapshots of the horizontal component of velocity $v_x(km/s)$ at times $t=50,150,250$ s in the $xz$ and $yz$ planes.}
  \end{figure*}

\begin{figure*}
  \begin{tabular}{ccc} \hline
  $t=50$ s & $t=150$ s & $t=250$ s \\ \hline
  && \\ 
 \multicolumn{3}{c}{$v_x(km/s)$ in $xy$ plane}\\ 
     \includegraphics[width=6.0cm]{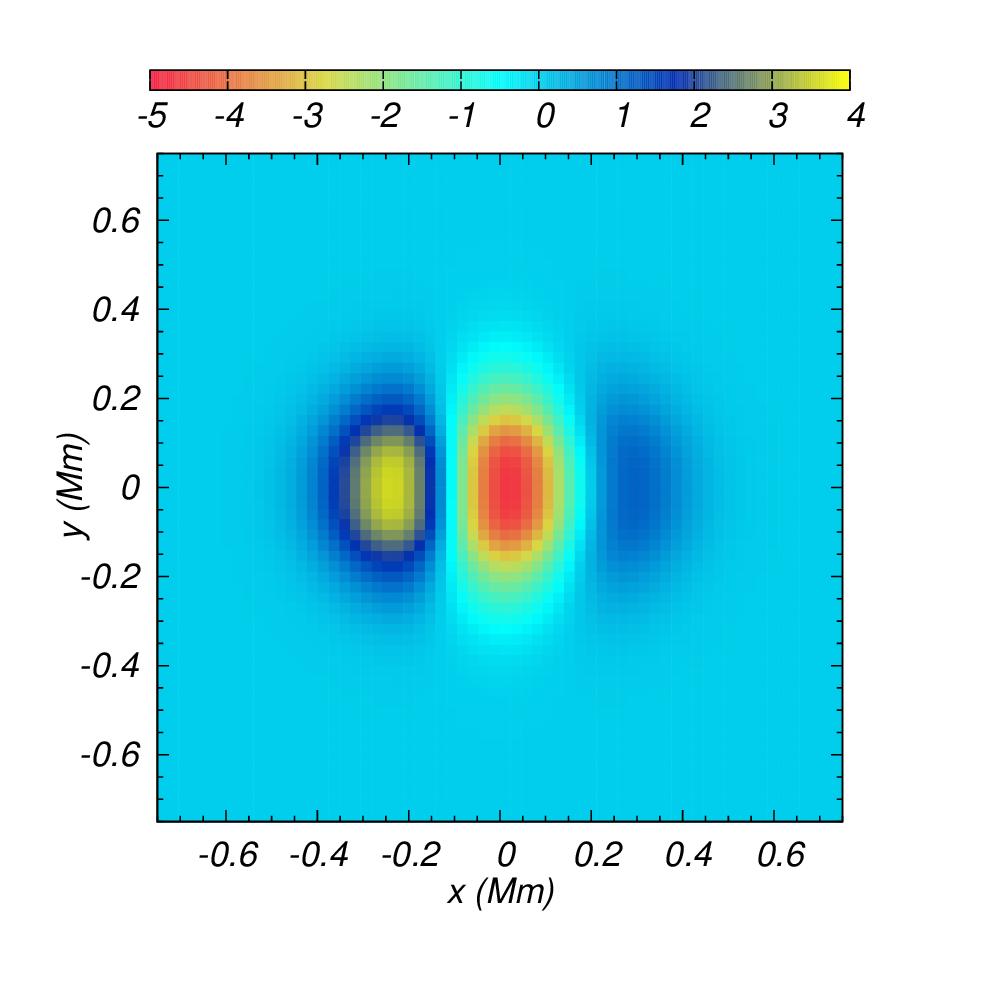}&
     \includegraphics[width=6.0cm]{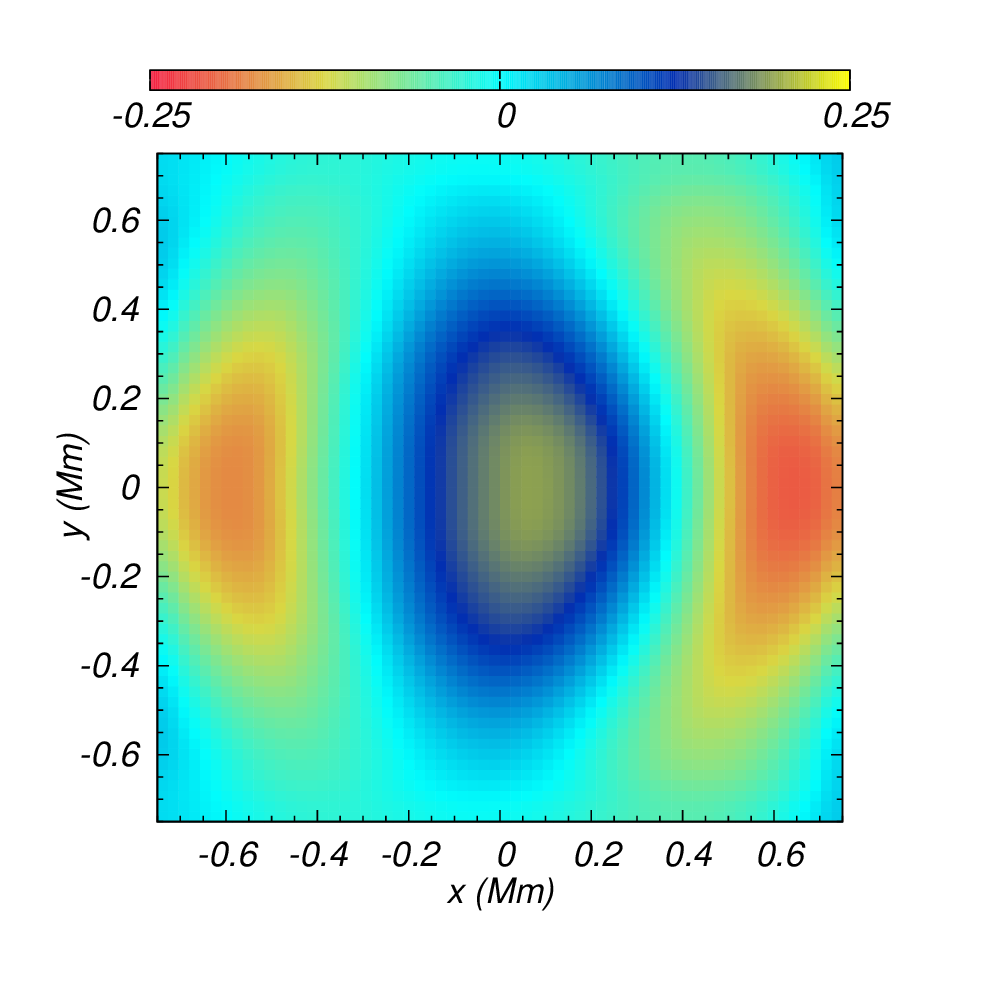}&
     \includegraphics[width=6.0cm]{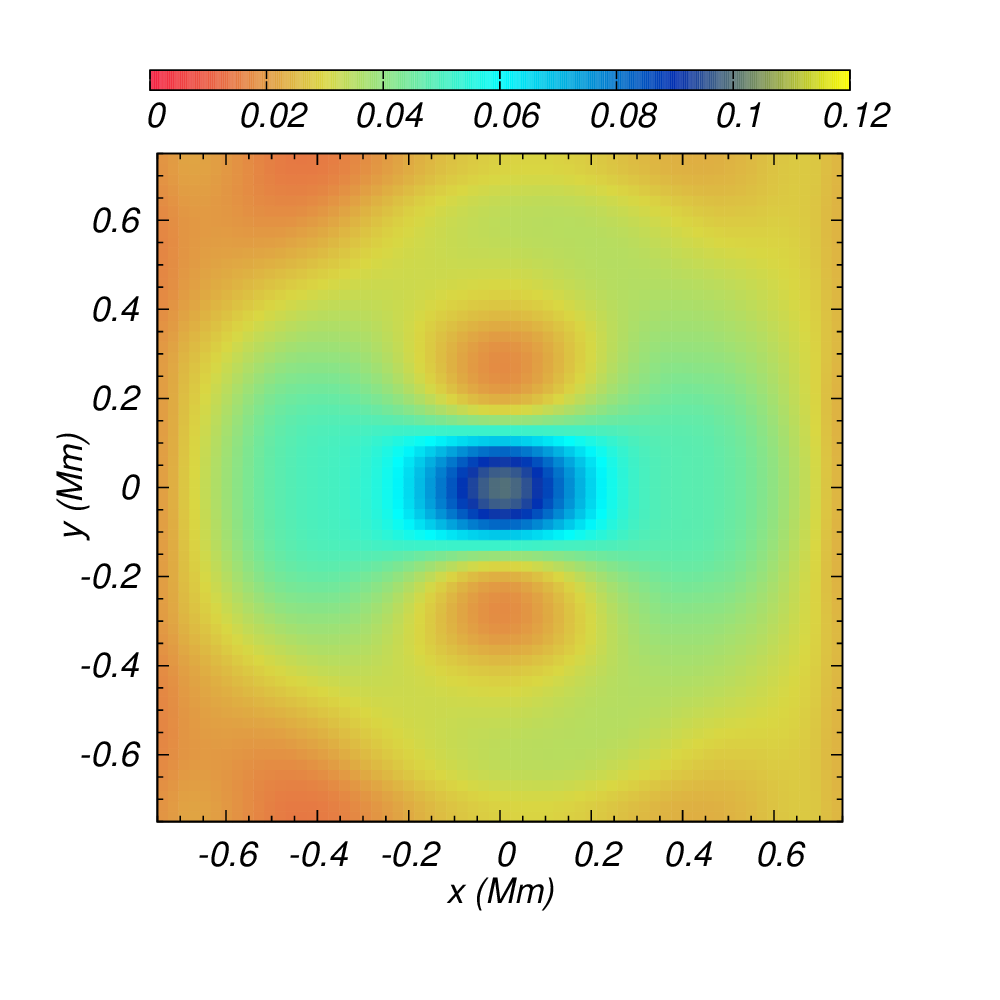}\\
   \multicolumn{3}{c}{$|\nabla\cdot B|(Tesla/km)$ in $xy$ plane}\\
     &
     \includegraphics[width=6.0cm]{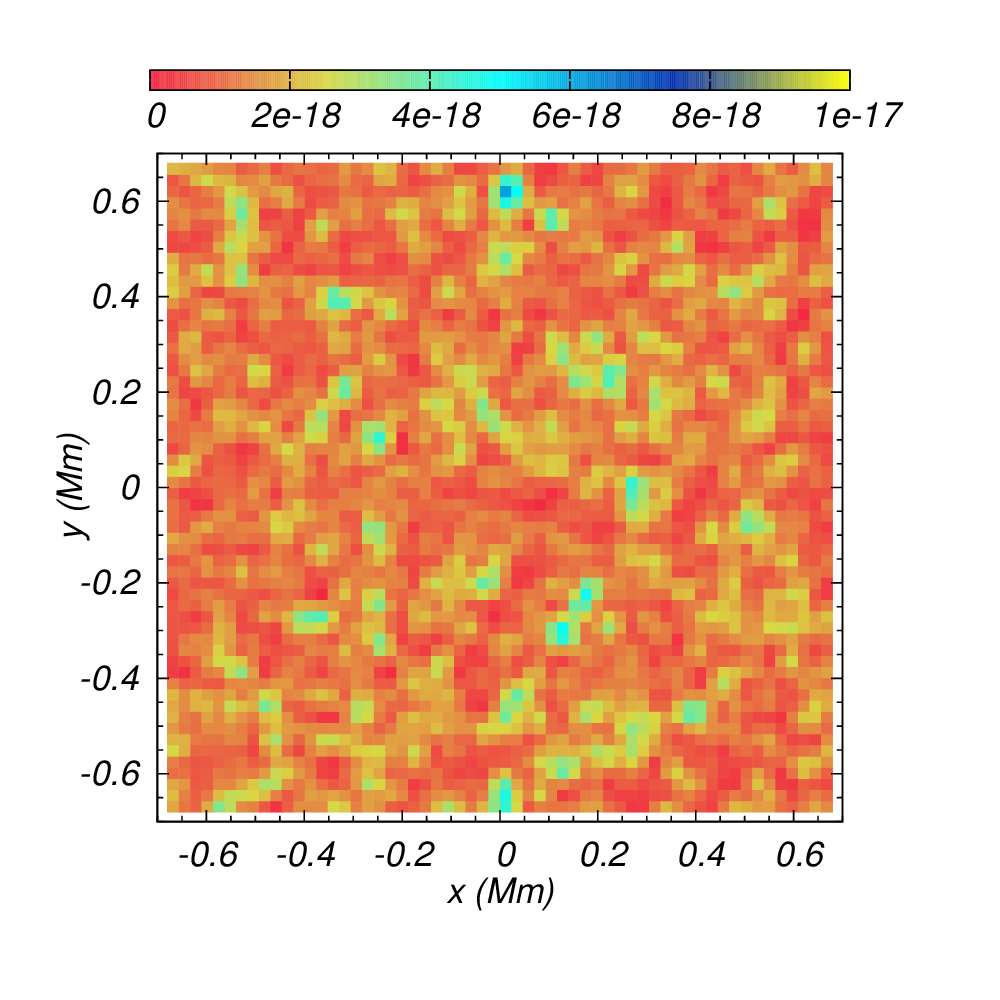}\\ 
 \hline
   \end{tabular}
   \caption{\label{fig:horizontal_perturbation_xy} Snapshots of the horizontal component of velocity $v_x(km/s)$ at times $t=50,150,250$ s and $|\nabla\cdot B|$ at $t=250$ s in the  $xy$ plane.}
  \end{figure*}

Horizontal perturbations in the solar atmosphere can be generated due to granular cells as a result of convection toward the photosphere. In this case we simulate these waves by exciting the $v_x$ component in equation (\ref{vx_perturbation}). This initial pulse excites all MHD waves which are affected by the gravitational constant field. Since early times we can see the magnetoacoustic-gravity waves propagating upwards in the  $xz$ plane as ligth yellow wave-fronts in Fig. \ref{fig:horizontal_perturbation_yz_xz} (top). At time $t=150$ s we can see the magnetoacoustic-gravity waves light yellow patches reaching the level $z=1.75$ Mm as shown in the top of Fig. 6 of \cite{Murawski_et_al_2013}. At time $t=250$ s, the waves reach the transition region, which causes the appareance of a light blue patch in Fig. \ref{fig:horizontal_perturbation_yz_xz}, but later on, they are  reflected back. In order to see the propagation of the waves from different planes, we show $v_x$ projected on the $xz$ and $yz$ planes in Fig. \ref{fig:horizontal_perturbation_yz_xz} (bottom). Unlike the previous test, in this case the velocity is not symmetrical. Moreover, the propagation of the waves seen from above show how the wave-fronts change their amplitudes until the reflection happens, this is shown in the top panel of Fig. \ref{fig:horizontal_perturbation_xy}. As in the previous case we show a snapshot of $|\nabla\cdot B|$ at time $t=150$ s in Fig. \ref{fig:horizontal_perturbation_xy} (bottom).

\begin{figure}
\begin{center}
\includegraphics[width=6.0cm]{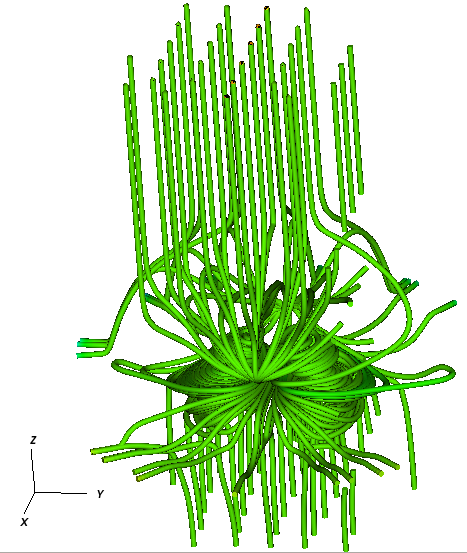}
\includegraphics[width=6.0cm]{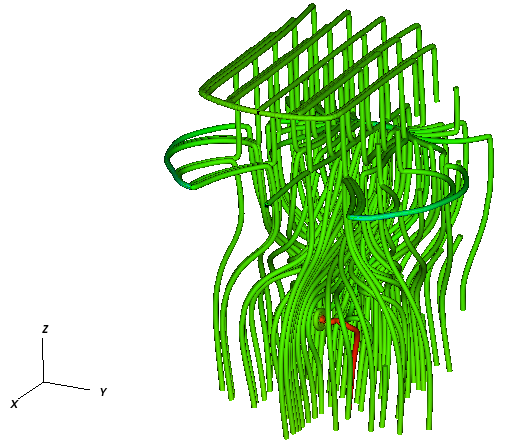}
\caption{\label{fig:streamlines_horizontal_pert} Temporal snapshots of streamlines at $t=50$ s (top) and $t=150$ s (bottom) for the case of the horizontal perturbation. Color represents the vorticity magnitude.}
\end{center}
\end{figure}

In this case we also show snapshots of the fluid streamlines at times $t=50$ s and $t=150$ s in Fig. \ref{fig:streamlines_horizontal_pert}. In the same way as in the case of the vertical perturbation, the horizontal perturbation generates vortices, but in this case the vorticity is more complex. It has been proposed that vortices are triggered by convective or granular motions \cite{Murawski_et_al_2013}.

% ------------------     SECTION     ----------------------
\section{Final comments}
\label{sec:conclusions}

We have presented a new  code that solves the ideal MHD equations intended to model phenomena in the solar atmosphere.

We have included standard 1D Riemann problem tests and shown that the order of convergence lies within the expected range corresponding to the methods we use. We conclude from our analysis, that MC and WENO5 perform better than the Minmod limiter.

We have also presented standards 2D tests including the Orszag-Tang vortex and a current sheet test problem showing magnetic reconnection. In these tests we have kept the magnetic field divergence free constraint under control, in all the tests near round-off error. We notice that, according to previous experience, due to the dissipation,  Minmod performs well and the MC and WENO5 limiters allow more refined structures in general.

The solar tests have been oriented to study wave propagation in a gravitationally stratified solar atmosphere. We track the transverse oscillations of Alfv\'enic pulses in solar coronal loops in a 2.5D model, in adittion we study the propagation of MHD-gravity waves and vortices on a 3D solar atmosphere. These results were compared to previous results in the literature. We have also shown that the constraint has been kept under control.

We have decided to use unigrid mode instead of mesh refinement in order to avoid the use of artificial viscosity at boundaries between refinement levels, which is used to damp out spurious oscillations. We also want to avoid the carbuncle effect at corners of refined grids. In order to carry out production runs with high resolution, we have parallelized our code.

Finally, the code is public and available under request at http://www.ifm.umich.mx/\~{}cafe

% ------------------------------------------------------------

% ----->     ACKNOWLEDGMENTS     <-----

\section*{Acknowledgments}

This research is partly supported by grant CIC-UMSNH-4.9. A. C-O gratefully acknowledges a DGAPA postdoctoral grant to Universidad Nacional Aut\'onoma de M\'exico (UNAM)

% -------------------------------------------------------
% -----     REFERENCES     ----------
% -------------------------------------------------------

\bsp

\label{lastpage}

\end{document}